\documentclass[sn-mathphys]{sn-jnl}

\usepackage{url}

\jyear{2022}%

\theoremstyle{thmstyleone}%
\theoremstyle{thmstyletwo}%

\theoremstyle{thmstylethree}%
\newtheorem{definition}{Definition}%

\begin{document}

\title[ ]{Non-linear dynamics of multibody systems: a system-based approach}


\author*[1]{\fnm{Daniel} \sur{Alazard}}\email{daniel.alazard@isae-supaero.fr}

\author[1]{\fnm{Francesco} \sur{Sanfedino}}\email{francesco.sanfedino@isae-supaero.fr}

\author[2]{\fnm{Ervan} \sur{Kassarian}}\email{ervan.kassarian@dycsyt.com}


\affil*[1]{\orgdiv{Féderation ENAC ISAE-SUPAERO ONERA, Université de Toulouse}, \orgaddress{\street{10 Avenue Marc Pélegrin}, \city{Toulouse}, \postcode{31400}, \country{France}}}
\affil[2]{\orgdiv{DyCSyT},  \orgaddress{\street{10 Avenue Edouard Belin}, \city{Toulouse}, \postcode{31400}, \country{France}}}
%


\abstract{
This paper presents causal block-diagram models to represent the equations of motion of multi-body systems in a very compact and simple closed form. 
Both the forward dynamics (from the forces and torques imposed at the various degrees-of-freedom to the motions of these degrees-of-freedom) or the inverse dynamics (from the motions imposed at the degrees-of-freedom to the resulting forces and torques) can be considered and described by a block diagram model. 
This work extends the Two-Input Two-Output Port (TITOP) theory by including all non-linear terms and uniform or gravitational acceleration fields. Connection among different blocks is possible through the definition of the motion vector.
The model of a system composed of a floating base, rigid bodies,  revolute and prismatic joints,  working under gravity is developed to illustrate the methodology. The proposed model is validated by simulation and cross-checking with a model built using an alternative modeling tool on a scenario where the nonlinear terms are determining.}

\maketitle
\section*{Nomenclature}
A calligraphic letter (for example $\mathcal{B}$) is used to label a rigid body. The same uppercase letter ($B$) is used to denote its center of mass. The subscript with the same lower case letter ($b$) is used to denote its body frame and reference axes (for example, $\mathcal{R}_b = (O,\mathbf{x}_b,\mathbf{y}_b,\mathbf{z}_b)$ is the body frame attached to $\mathcal B$ at the reference point $O$ ). $\mathcal{R}_i = (I,\mathbf{x}_i,\mathbf{y}_i,\mathbf{z}_i)$ is the inertial frame. In addition, the following notations will be used throughout this paper:


\hspace{-0.75cm}\begin{tabular}{{l}p{10.cm}}
	$\mathbf{P}_{a/b}$ & $3 \times 3$ Direction Cosine Matrix (DCM) of the rotation from frame $\mathcal{R}_a$ to frame $\mathcal{R}_b$, that contains the coordinates of vectors $\mathbf{x}_a$, $\mathbf{y}_a$, $\mathbf{z}_a$ expressed in frame $\mathcal{R}_b$. \\
		$\overrightarrow{IP}$ & The vector from point $I$ to point $P$ ($3 \times 1$ vector, $m$).\\
	$\boldsymbol{\Theta}^\mathcal{B}$ & The \textsc{Euler} angle vector of the attitude of the frame $\mathcal R_b$  w.r.t. $\mathcal R_i$ for a given rotation sequence ($3 \times 1$ vector, $rad$)\\
	$\mathbf P_{./i}(\boldsymbol \Theta^\mathcal{B})$ & The \textsc{Euler} angles to DCM conversion function (for the given rotation sequence): $\mathbf{P}_{b/i}=\mathbf P_{./i}(\boldsymbol \Theta^\mathcal{B})$.\\
	$\boldsymbol\Theta(\mathbf{P}_{b/i})$ & The inverse function : $\boldsymbol{\Theta}^\mathcal{B}=\boldsymbol\Theta(\mathbf{P}_{b/i})$.\\
	$\mathbf{v}^\mathcal{B}_P$ & Inertial velocity of body $\mathcal{B}$ at point $P$ ($3 \times 1$ vector, $m/s$).\\
	$\boldsymbol{\omega}^\mathcal{B}$ & Angular speed of $\mathcal{R}_b$ with respect to  $\mathcal{R}_i$ ($3 \times 1$ vector, $rad/s$).\\
	$\mathbf{a}^\mathcal{B}_P$ & Inertial acceleration of body $\mathcal{B}$ at point $P$ ($3 \times 1$ vector, $m/s^2$).\\
	$\mathbf{F}_{\mathcal{B/A}}$ & Force applied by body $\mathcal{B}$ on body $\mathcal{A}$ ($3 \times 1$ vector, $N$).\\
	$\mathbf{T}_{\mathcal{B/A},P}$ & Torque applied by $\mathcal{B}$ on $\mathcal{A}$ at point $P$ ($3 \times 1$ vector, $Nm$).\\
	$\mathbf{F}_{ext/\mathcal B}$  & Total external force applied to body $\mathcal{B}$ ($3 \times 1$ vector, $N$).\\
    $\mathbf{T}_{ext/\mathcal B,P}$ & Total external torque applied to body $\mathcal{B}$ at the point $P$ ($3 \times 1$ vector, $Nm$).\\
	$\left[\star\right]_{\mathcal R_b}$ & Projection of $\star$ (vector, wrench, tensor, model,...) in the frame $\mathcal R_b$. For a  vector $\mathbf v$: $[\mathbf v]_{\mathcal R_b}=\mathbf{P}_{a/b}[\mathbf v]_{\mathcal R_a}$. \\
	$\mathbf{W}_{\mathcal{B/A},P}$ & $6 \times 1$  wrench applied by $\mathcal{B}$ on $\mathcal{A}$ at point $P$: $\mathbf{W}_{\mathcal{B/A},P}=\left[\begin{array}{c}\mathbf{F}_{\mathcal{B/A}}\\ \mathbf{T}_{\mathcal{B/A},P}\end{array}\right]$.\\
	$\mathbf{W}_{ext/\mathcal B,P}$ & $6 \times 1$ total external wrench applied on $\mathcal{B}$  at point $P$.
	\\
	$\mathbf{W}_{\mathcal{./A},P}$ & $6 \times 1$  local wrench applied on $\mathcal{A}$ at point $P$.
	\\
	$\mathbf{W}_{\mathcal{A/.},P}$ & $6 \times 1$  local wrench applied by $\mathcal{A}$ at point $P$.\\
	${\left.\frac{d \mathbf{v}}{dt}\right\vert }_{\mathcal{R}}$ & Time-derivative w.r.t. frame $\mathcal{R}$ of the vector $\mathbf{v}$ $\left(\left[\frac{d \mathbf{v}}{dt}\vert_{\mathcal{R}}\right]_{\mathcal R}=\frac{d\left[ \mathbf v\right]_{\mathcal{R}}}{dt}\right)$ .\\
	$\mathbf{\dot v}^{\mathcal B}$ & Time-derivative of the vector $\mathbf v^{\mathcal B}$ in the body frame: $\mathbf{\dot v}^{\mathcal B}={\frac{d \mathbf{v}^{\mathcal B}}{dt}\vert}_{\mathcal{R}_b}$\\
	$\mathbf{x}^\mathcal{B}_P$ & $6 \times 1$ pose dual vector of body $\mathcal B$ at point $P$: $\mathbf{ x}^\mathcal{B}_P=\left[\begin{array}{c} \overrightarrow{IP} \\ \boldsymbol{\Theta}^\mathcal{B}\end{array}\right]$.\\
	$\boldsymbol{\mathbf x'}^\mathcal{B}_P$ & $6 \times 1$ twist (dual vector) of body $\mathcal B$ at point $P$: $\mathbf{ x'}^\mathcal{B}_P=\left[\begin{array}{c}\mathbf{v}^\mathcal{B}_P \\ \boldsymbol{\omega}^\mathcal{B}\end{array}\right]$.\\ 
	$\boldsymbol{\mathbf x''}^\mathcal{B}_P$ & $6 \times 1$  acceleration dual vector of body $\mathcal B$ at point $P$: $\mathbf{ x''}^\mathcal{B}_P=\left[\begin{array}{c}\mathbf{a}^\mathcal{B}_P \\ \boldsymbol{\dot\omega}^\mathcal{B}\end{array}\right]$.\\
	$\mathbf m^\mathcal{B}_P$ & The $18$ components motion vector of body $\mathcal B$ at point $P$; for the purposes of notations: $\mathbf m^\mathcal{B}_P=\left[{\boldsymbol{\mathbf{\dot x}'}^\mathcal{B}_P}^T\;\;{\boldsymbol{\mathbf x'}^\mathcal{B}_P}^T\;\;{\mathbf{x}^\mathcal{B}_P}^T\right]^T$ 
	\end{tabular}

\hspace{-0.75cm}\begin{tabular}{{l}p{10.cm}}

$\boldsymbol{\Gamma(\Theta)}$ & The relation from angular velocity to  \textsc{Euler} angle time-derivatives (for the given rotation sequence) $\dot{\boldsymbol\Theta}^{\mathcal B}=\boldsymbol\Gamma(\boldsymbol\Theta^\mathcal{B})[\boldsymbol\omega^\mathcal{B}]_{\mathcal R_b}$.	\\
$m^{\mathcal{B}}$ &  mass of body $\mathcal{B}$.\\
$\mathbf{I}^{\mathcal{B}}_P$ &  $3 \times 3$  inertia tensor of body $\mathcal{B}$ at point $P$.\\
$\mathbf{D}^{\mathcal{B}}_P$ & $6\times 6$ static dynamic model of body $\mathcal B$ expressed at point $P$.\\
$\boldsymbol{\tau}_{PB}$ &  Kinematic model between points $P$ and $B$:\\
& 
	$\boldsymbol{\tau}_{PB}=\left[\begin{array}{cc}
		\mathbf{1}_3  & (^*\overrightarrow{PB})\\
		\mathbf{0}_{3\times 3} & \mathbf{1}_3  \\
	\end{array}\right]$.\\
	$(^*\mathbf v)$ & Skew symmetric matrix associated with vector $\mathbf v$: if $\left[\mathbf v\right]_{\mathcal{R}}=\left[\begin{array}{c}x \\y \\ z\end{array}\right]$ then $\left[(^*\mathbf v)\right]_{\mathcal{R}}=  \left[\begin{array}{ccc}
		0  & -z & y\\
		z & 0 & -x  \\
		-y & x & 0
	\end{array}\right]$.\\
	$\mathbf{g}$ &  gravitational or uniform acceleration ($3 \times 1$ vector, $m/s^2$).\\
	$\mathbf{r},\;\mathbf t$ & directions of the revolute, prismatic joint axes ($3\times 1$ vectors, $m$).\\
	$\mathbf{\tilde r},\;\mathbf{\tilde t}$ & unit vectors along $\mathbf{ r},\;\mathbf{t}$ ($3\times 1$ vectors).\\
	$\theta,\;\theta_0$ & angular configuration, initial value of the revolute joint ($rad$). \\
	$x,\;x_0$ & linear configuration, initial value of the prismatic joint ($m$). \\
	$\mathrm{s}$ &  \textsc{Laplace}'s variable.\\
	$\mathbf{1}_n$ &  Identity matrix $n \times n$.\\
	$\mathbf{0}_{n\times m}$  & Zero matrix $n \times m$.\\
	$\mathbf{A}^T$ & Transpose of $\mathbf{A}$.\\
	$\mathrm{diag}$ & Diagonal augmentation.\\
	$\mathbf{P}_{a/b}^{\times 2}$ & Augmented DCM for dual vectors $\mathbf{P}_{a/b}^{\times 2}=\mathrm{diag}\left(\mathbf{P}_{a/b},\,\mathbf{P}_{a/b}\right)$.\\
	$\mathbf P_{a/b}^{18}$ &  Augmented DCM for motion vectors: \\ & $\mathbf P_{a/b}^{18}=\mathrm{diag}\left(\mathbf{P}_{a/b}^{\times 2},\;\mathbf{P}_{a/b}^{\times 2},\;\mathbf{P}_{a/b},\;\mathbf 1_3\right)$.
\end{tabular}
The reader will find in \cite{Schaub} (Chapter 3) the expressions of $\mathbf P_{./i}(\boldsymbol \Theta^\mathcal{B})$,  $\boldsymbol\Theta(\mathbf{P}_{b/i})$, $\boldsymbol{\Gamma(\Theta)}$  and the basic background in kinematics and attitude parameterization.

\section{Introduction}
The modeling of rigid Multi-Body System (MBS) have motivated lots of contributions during the last decades (\cite{Hahn,Roy}), more particularly on the methods to derive the equations of motions and the algorithms to perform efficient simulations from the computational cost point of view (\cite{Shabana,Simeon}). Focusing on the equations of motions which is the main concern of this paper, rather than their numerical implementation, one can distinguish the methods based on the virtual work principle (\textsc{Euler-Lagrange equations}) which can provide a reduced set of  differential equations under the form:
\begin{equation}
	\mathbf M(\mathbf q)\ddot{\mathbf q}+\mathbf C(\mathbf q, \dot{\mathbf q})=\boldsymbol \tau
\end{equation}
where $\mathbf q$ is the vector of the generalized coordinates or independent degrees-of-freedom (d.o.f), $\mathbf M(\mathbf q)$ is the mass matrix (also called the direct dynamic model when only  the linear behavior is considered as proposed in \cite{oatao16559}), $\mathbf C(\mathbf q, \dot{\mathbf q})$ are the Coriolis and centrifugal terms and $\boldsymbol \tau$ is the vector of generalized forces. These equations do not include the constraint forces due to the joints between the bodies locking some relative d.o.f. (also called internal forces or joint forces). They can be augmented by an algebraic equation involving \textsc{Lagrange} multipliers to take into loop closure constraints in the case of an MBS with closed kinematic chains. The model thus obtained is governed by a set of differential-algebraic equations. One can also distinguish the \textsc{Newton-Euler}-based methods \cite{Roy} involving redundant coordinates associated to the individual rigid bodies and additional motion constraints due to the joints. In both cases, the obtained equations of motions are used to perform two different types of analysis, namely the inverse dynamics and the forward dynamics. The first one aims to compute the required forces to produce a given motion (or trajectory in terms of accelerations $\ddot{\mathbf q}$, velocities $ \dot{\mathbf q}$ and position $\mathbf q$ of the d.o.f ). This analysis requires to solve only algebraic equations. The second one (forward dynamics) aims to compute the accelerations and then the velocities and positions by integration from the forces imposed at the d.o.fs.  In most applications, a closed-form of the equations of motion cannot be obtained and one must use algorithms based on several (forward and backward) recurrences over the open kinematic chain or tree of bodies to compute the mass matrix $\mathbf M(\mathbf q)$, the coriolis and centrifugal terms $\mathbf C(\mathbf q, \dot{\mathbf q})$ or to solve the inverse or forward dynamics. One can mention  the Recurrent \textsc{Newton-Euler} Algorithm (RNEA) to solve the inverse dynamics or the Articulated-Body Algorithm (ABA) or the Decoupled Natural Orthogonal Complement (DeNOC, \cite{DeNOC,Vincent}) to solve the forward dynamics. These algorithms proceed to several recurrence over the bodies of the MBS ($2$ passes for the inverse dynamics and $3$ passes for the forward dynamics)  and, although they are efficient from the computational point of view, they are not adapted to represent simply and straightforwardly the equations of motions.

The main contributions of this paper is to propose a block-diagram model to represent the equations of motions for arbitrary open or closed kinematics chains or trees of rigid bodies. This block-diagram model involves as many blocks as bodies and connecting wires to propagate the wrenches and the motions (accelerations, velocities and positions). Both the inverse dynamics and the forward dynamics can be solved by a dedicated block-diagram model. The  forward and backward recurrent passes required in the mentioned algorithms are simply taken into account  by an algebraic loop created by the interconnections between the blocks (or bodies). The main interest of such a block-diagram description of the MBS is its modularity allowing an update in a particular body or subsystem to be taken into account very easily and allowing fast prototyping. One can mention Simscape-Multibody toolbox \cite{mathworks2021simscape} as an example of user-friendly block-diagram modeling tool for MBS. Note that in  Simscape-Multibody the block-diagrams are acausal and the way that the equations of motions are generated is not detailed. In the proposed approach, the block-diagrams are causal and different for the inverse dynamics and the forward dynamics. All the equations of motions, based on revisited \textsc{Newton-Euler} equations and required to build the various blocks, are detailed in the following sections. This paper should be seen as the non-linear generalization of the Two-Input Two-Output Port (TITOP) modeling approach presented in  \cite{oatao16559} and \cite{AlazardMUBO2023} but restricted, till now, to Linear Parameter Varying (LPV) models of  rigid or flexible MBS. The interest of TITOP  LPV models and the associated Satellite Dynamics Toolbox Library (SDTlib) \cite{alazard2020} to assess robust pointing performances in space engineering  or to perform mechanical/ control co-design is highlighted in \cite{SANFEDINO2022107961} and \cite{Finozzi2022}, respectively. Note also that the TITOP approach presented in   \cite{oatao16559}  provides a linear model valid for small motions around null equilibrium conditions and thus cannot take into account parameter-dependent equilibrium conditions. In some particular applications, dedicated linear models can be developed as proposed in \cite{kassarianIEEE} for a ballon-borne telescope subjected to Earth gravity and subject to varying mass or in \cite{rodrigues2024modelinganalysisflexiblespinning} where the equilibrium conditions (centrifugal loads) depend on the varying spin rate of the MBS. The work presented in this paper is restricted to rigid MBS but captures all the non-linear terms. It should be considered as an intermediate result in a longer-term development aiming at creating a rigid/flexible MBD modeling tool allowing to derive LPV models around any equilibrium condition for parametric robustness analysis and control and also enabling high-fidelity nonlinear simulations for validation purposes.

The proposed generalization of the TITOP approach is based on the  \textsc{Newton-Euler} equations reformulated in terms of spatial acceleration \cite{Roy} in Section \ref{sect:NE}. It will be shown that the nonlinear TITOP model of a rigid body, detailed in Section \ref{sec:TITOP}, includes in the same model the forward dynamics at the connecting point with a child body and the inverse dynamics at the connecting point with the parent body. The nonlinear TITOP models for bodies connected through a revolute or a prismatic joint are also detailed as well as a loop closure block to cope with MBS with closed-kinematics chain. In the last section, the model of a system composed of a floating base, rigid bodies,  revolute and prismatic joints,  working under gravity is developed to illustrate the methodology. The proposed model is validated by simulation and cross-checking with a model built using Simscape/Multibody on a scenario where the nonlinear terms are determining.

\section{Revisited \textsc{Newton-Euler} equation}\label{sect:NE}
The \textsc{Newton-Euler} equation applied to a rigid body $\mathcal B$ at the point $P$, distinct from its center of mass $B$ reads \cite{wiki:NE}:
\begin{eqnarray}\nonumber
	\left[\begin{array}{c}\mathbf{F}_{ext/\mathcal B}\\ \mathbf{T}_{ext/\mathcal B,P}\end{array}\right]&=&
	\underbrace{\left[\begin{array}{cc}m^{\mathcal{B}}\mathbf{1}_3 & m^{\mathcal{B}}\,(^*\overrightarrow{BP})\\ -m^{\mathcal{B}}\,(^*\overrightarrow{BP}) & \mathbf{I}^{\mathcal{B}}_B -m^{\mathcal{B}}\,(^*\overrightarrow{BP})^2\end{array}\right]}
	_{\mathbf{D}^{\mathcal{B}}_P}
	\left[\begin{array}{c} \mathbf{a}^{\mathcal B}_P\\  \boldsymbol{\dot\omega}^{\mathcal B}\end{array}\right]\cdots \\ \label{eq:NEfull} & & \cdots +
	\underbrace{\left[\begin{array}{c}m^{\mathcal{B}} (^*\boldsymbol \omega^{\mathcal B})(^*\overrightarrow{BP})\boldsymbol\omega^{\mathcal B}  \\ (^*\boldsymbol\omega^{\mathcal B})\left(\mathbf{I}^{\mathcal{B}}_B -m^{\mathcal{B}}\,(^*\overrightarrow{BP})^2 \right)\boldsymbol\omega^{\mathcal B}\end{array}\right]}_{\mathbf{W}^{\mathcal B}_P(\boldsymbol\omega^{\mathcal B})}
\end{eqnarray}
or using the notation defined in the nomenclature:
\begin{equation}
	\mathbf{W}_{ext/\mathcal B,P}=\mathbf{D}^{\mathcal{B}}_P \boldsymbol{\mathbf x''}^\mathcal{B}_P+ \mathbf{W}^{\mathcal B}_P(\boldsymbol\omega^{\mathcal B})
\end{equation}
This equation is used to described the 6 degrees-of-freedom (d.o.f.) motion of a rigid body free to move in space while taking into account all the couplings between the 3 translations and the 3 rotations which appear when $P\neq B$ (see also: \cite{Hahn}). The first element of the right-hand term describes the linear model between the  acceleration dual vector $ \mathbf{x''}^\mathcal{B}_P$ and the applied resultant external wrench $\mathbf{W}_{ext/\mathcal B,P}$ through the $6\times 6$ dynamic model $\mathbf{D}^{\mathcal{B}}_P$ of the body $\mathcal B$ at the point $P$. This dynamic model can be easily expressed from the mass $m^\mathcal B$, the $3\times 3 $ inertia tensor $\mathbf I^\mathcal B_B$ at the center of mass $B$ and the kinematic model $\boldsymbol \tau_{BP}$ between the points $B$ and $P$:
\begin{equation}\mathbf{D}^{\mathcal{B}}_P=\boldsymbol\tau_{BP}^T \underbrace{\left[\begin{array}{cc}m^{\mathcal{B}}\mathbf{1}_3 &\mathbf{0}_{3\times 3}\\\mathbf{0}_{3\times 3} & \mathbf{I}^{\mathcal{B}}_B \end{array}\right]}_{\mathbf{D}^{\mathcal{B}}_B}\boldsymbol\tau_{BP}
	\end{equation}
The second element $\mathbf{W}^{\mathcal B}_P(\boldsymbol\omega^{\mathcal B})$ is the wrench due to the non-linear terms also called the terms in "$\omega$-squared", usually neglected when the angular rates are small. Note also that the equation \eqref{eq:NEfull} is valid in any frame but is commonly projected in the body frame $\mathcal R_b$ in which the model $\mathbf{D}^{\mathcal{B}}_P$ is time-invariant.

By introducing the \textbf{spatial acceleration} or the \textbf{acceleration w.r.t to the body frame}
$\mathbf{\dot v}^{\mathcal B}_P$ such that:

\[
\mathbf{a}^{\mathcal B}_P=\left.\frac{d \mathbf{v}^{\mathcal B}_P}{dt}\right\vert_{\mathcal{R}_b}+(^*\boldsymbol\omega^{\mathcal B})\mathbf{v}^{\mathcal B}_P=\mathbf{\dot v}^{\mathcal B}_P+(^*\boldsymbol\omega^{\mathcal B})\mathbf{v}^{\mathcal B}_P\;
\]
then \textsc{Newton-Euler} equations reads:
\begin{eqnarray}\nonumber
	\mathbf{W}_{ext/\mathcal B,P}=\left[\begin{array}{c}
		\mathbf{F}_{ext}\\
		\mathbf{T}_{ext,P}
	\end{array} \right] &= &\mathbf{D}_P^{\mathcal B} \left[\begin{array}{c}
		\mathbf{\dot v}^{\mathcal B}_P\\
		\boldsymbol{\dot\omega}^{\mathcal B}
	\end{array}\right] + \left[ \begin{array}{c}
		m^{\mathcal B}(^*\boldsymbol{\omega}^{\mathcal B}) \mathbf{v}^{\mathcal B}_P\\
		- m^{\mathcal B}(^*\overrightarrow {BP} )(^*\boldsymbol{\omega}^{\mathcal B}) \mathbf{v}^{\mathcal B}_P
	\end{array} \right] \cdots \\ \label{eq:NEetPinRp}  & & \cdots + \left[\begin{array}{c}
		m^{\mathcal B}(^*\boldsymbol{\omega}^{\mathcal B})(^*\overrightarrow {BP} )\boldsymbol{\omega}^{\mathcal B}\\
		(^*\boldsymbol{\omega}^{\mathcal B})\left(\mathbf{I}_B^{\mathcal B} - m^{\mathcal B}(^*\overrightarrow {BP} )^2 \right)\boldsymbol{\omega}^{\mathcal B}
	\end{array} \right]\;.
\end{eqnarray}
Using the \textsc{Jacobi} identity:
\begin{equation}
	(^*\overrightarrow {BP})(^ *\boldsymbol{\omega}^{\mathcal B})\bf{v}^{\mathcal B}_P = (^ * \boldsymbol{\omega}^{\mathcal B})(^*\overrightarrow {BP} )\bf{v}^{\mathcal B}_{P} - (^ * \bf{v}^{\mathcal B}_{P})(^*\overrightarrow {BP} )\boldsymbol{\omega}^{\mathcal B}
\end{equation}
and considering that $(^*\bf{v}^{\mathcal B}_{P})\mathbf{v}^{\mathcal B}_P=0$, then equation \eqref{eq:NEetPinRp} can be factorized under the very compact form
 :
\begin{equation}\label{eq:NEetPinRp_f}
	\mathbf{W}_{ext/\mathcal B,P}=\left[ {\begin{array}{c}
			\mathbf{F}_{ext}\\
			\mathbf{T}_{ext,P}
	\end{array}} \right]= \bf{D}_P^{\mathcal B} \left[\begin{array}{c}
		\mathbf{\dot v}^{\mathcal B}_P\\
		\boldsymbol{\dot{\omega}}^{\mathcal B}
	\end{array} \right] + \underbrace{\left[\begin{array}{cc}
		(^ *\boldsymbol{\omega}^{\mathcal B})&{\bf{0}}_3\\
		(^ *\bf{v}^{\mathcal B}_P)&(^ * \boldsymbol{\omega} ^{\mathcal B})
	\end{array} \right]}_{\boldsymbol{\mathcal C}(\boldsymbol{\mathbf x'}^{\mathcal B}_P)}\bf{D}_P^{\mathcal B} \left[ \begin{array}{c}
		\bf{v}^{\mathcal B}_P\\
		\boldsymbol{\omega}^{\mathcal B} 
	\end{array} \right]\;.
\end{equation}
\begin{equation}\label{eq:NE_short}
	\mbox{or: } \mathbf{W}_{ext/\mathcal B,P}=\mathbf{D}^{\mathcal{B}}_P \boldsymbol{\dot{\mathbf x}'}^\mathcal{B}_P+\boldsymbol{\mathcal C}(\boldsymbol{\mathbf x'}^{\mathcal B}_P)\mathbf{D}^{\mathcal{B}}_P \boldsymbol{\mathbf x'}^\mathcal{B}_P
\end{equation}
This formulation of the \textsc{Newton-Euler}, projected in the body frame $\mathcal R_b$,  is also linked to the \textsc{Lagrange} derivation with quasi-coordinates \cite{meirovitch1970methods}:
\begin{equation}\label{eq:Lagrane_quadsi_coordinates}
	\frac{d}{dt}\frac{\partial L^{\mathcal{B}}}{\partial\left[\begin{array}{c}\mathbf{v}^{\mathcal B}_P\\ \boldsymbol{\omega}^{\mathcal B}\end{array}\right]_{\mathcal{R}_b}}+ \left[\boldsymbol{\mathcal C}(\boldsymbol{\mathbf x'}^{\mathcal B}_P)\right]_{\mathcal{R}_b}\frac{\partial L^{\mathcal{B}}}{\partial\left[\begin{array}{c}\mathbf{v}^{\mathcal B}_P\\ \boldsymbol{\omega}^{\mathcal B}\end{array}\right]_{\mathcal{R}_b}}-\left[\begin{array}{cc}
		\mathbf{P}_{b/i}^T&\mathbf{0}_3\\ \mathbf{0}_3 &
		\boldsymbol{\Gamma}^T
	\end{array} \right]\frac{\partial L^{\mathcal{B}}}{\partial\left[\begin{array}{c}
			[\overrightarrow{IP}]_{\mathcal{R}_i} \\ \boldsymbol{\Theta}^{\mathcal B}	\end{array}\right]}=\left[\mathbf{W}_{ext/\mathcal B,P}\right]_{\mathcal{R}_b}
\end{equation}
where $L^{\mathcal B}$ is the Lagrangian of body $\mathcal B$. Indeed, in our case (a single rigid body in space), the potential energy is null and the Lagrangian is equal the the kinetic energy and does not depend on the position of the body $\overrightarrow{IP}$ or its attitude $\boldsymbol{\Theta}^\mathcal{B}$:
 \begin{equation}
 	L^{\mathcal{B}}=\frac{1}{2}\left[\begin{array}{c}\mathbf{v}^{\mathcal B}_P\\ \boldsymbol\omega^{\mathcal B}\end{array}\right]_{\mathcal{R}_b}^T\left[\mathbf{D}_P^{\mathcal{B}}\right]_{\mathcal{R}_b}\left[\begin{array}{c}\mathbf{v}^{\mathcal B}_P\\ \boldsymbol \omega^{\mathcal B}\end{array}\right]_{\mathcal{R}_b}=\frac{1}{2}\left[\boldsymbol{\mathbf x'}^{\mathcal B}_P\right]_{\mathcal{R}_b}^T\left[\mathbf{D}_P^{\mathcal{B}}\right]_{\mathcal{R}_b}\left[\boldsymbol{\mathbf x'}^{\mathcal B}_P\right]_{\mathcal{R}_b}\;.
\end{equation}
A gravitational or uniform acceleration $\mathbf g$ can also be taken into account in the following modified \textsc{Newton-Euler} equation:
\begin{equation}\label{eq:NE_short_g}
	\mathbf W_{ext/\mathcal{B},P}=\textbf{D}_P^{\mathcal B}\left( \boldsymbol{\dot{\mathbf x}'}^{\mathcal B}_P-\left[\begin{array}{c}\mathbf g \\ \mathbf 0_{3 \times 1}\end{array}\right]\right) + \boldsymbol{\mathcal C}(\mathbf {x'}^{\mathcal B}_P)\textbf{D}_P^{\mathcal B} 
		\boldsymbol{\mathbf x'}^{\mathcal B}_P\;.
\end{equation}
In the TITOP approach presented in \cite{oatao16559}, and restricted to the linear behavior of multi-body systems, only the acceleration was propagate in the block-diagram model. The  non-linear model \eqref{eq:NE_short} requires to propagated the spatial acceleration dual vector w.r.t. to the body frame $\boldsymbol{\dot{\mathbf x}'}^{\mathcal B}_P$ ($\neq\boldsymbol{\mathbf x''}^\mathcal{B}_P$)  and the twist $\mathbf {x'}^{\mathcal B}_P$. Considering a uniform gravitational acceleration in the inertial frame (i.e.:  $[\mathbf g]_{\mathcal R_i}=const$), then the projection of the model \eqref{eq:NE_short_g} in the body frame $\mathcal R_b$ requires also to propagate the attitude $\boldsymbol\Theta^\mathcal B$ of the body as proposed in \cite{kassarianIEEE}. Thus for non-linear TITOP models, the whole motion vector $\mathbf m^{\mathcal B}_P$ of the body $\mathcal B$ at the point $P$ will be propagated. As defined in the nomenclature this $18$ components motion vector gathers the spatial acceleration dual vector $\boldsymbol{\dot{\mathbf x}'}^{\mathcal B}_P$, the twist $\mathbf {x'}^{\mathcal B}_P$ and the pose dual vector (position and attitude) $\mathbf {x}^{\mathcal B}_P$ for the $6$ d.o.f.

The equation of motion \eqref{eq:NE_short_g}  can then be completed by the kinematics equations:
\begin{itemize}
	\item $\mathbf v_P^{\mathcal B}={\left.\frac{d \overrightarrow{IP}}{dt}\right\vert}_{\mathcal R_b}-(^*\overrightarrow{IP})\boldsymbol{\omega}^{\mathcal B}=\dot{\overrightarrow{IP}}-(^*\overrightarrow{IP})\boldsymbol{\omega}^{\mathcal B}$,
	\item $\boldsymbol{\dot\Theta}^{\mathcal B}=\boldsymbol{\Gamma}(\boldsymbol{\Theta}^{\mathcal B}) \left[ \boldsymbol{\omega}^{\mathcal B}\right]_{\mathcal{R}_b}$,
\end{itemize}
Thus:
\begin{equation}\label{eq:kin}
\mathbf {\dot x}^{\mathcal B}_P=
\left[\begin{array}{cc}\mathbf{1}_3 & (^*\overrightarrow{IP}) \\
	\mathbf{0}_{3\times 3} & \boldsymbol\Gamma(\boldsymbol{\Theta}^{\mathcal B})
\end{array}\right]\mathbf {x'}^{\mathcal B}_P\;.
\end{equation}

\begin{figure}[!ht]
		\begin{center}
		\includegraphics[angle=0, width=0.8\textwidth]{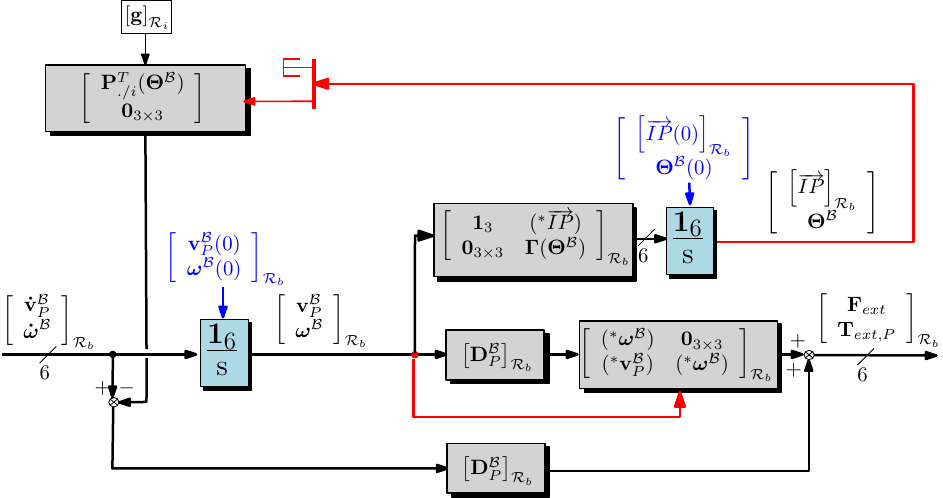}
		\caption{Block diagram representation of the direct dynamic model of the body $\mathcal B$ at the point $P$ and projected in $\mathcal R_b$
			.}
		\label{fig:NEinv_atP_inRb_MUBO}
			\end{center}
\end{figure}

\begin{figure}[!ht]
		\begin{center}
		\includegraphics[angle=0, width=\textwidth]{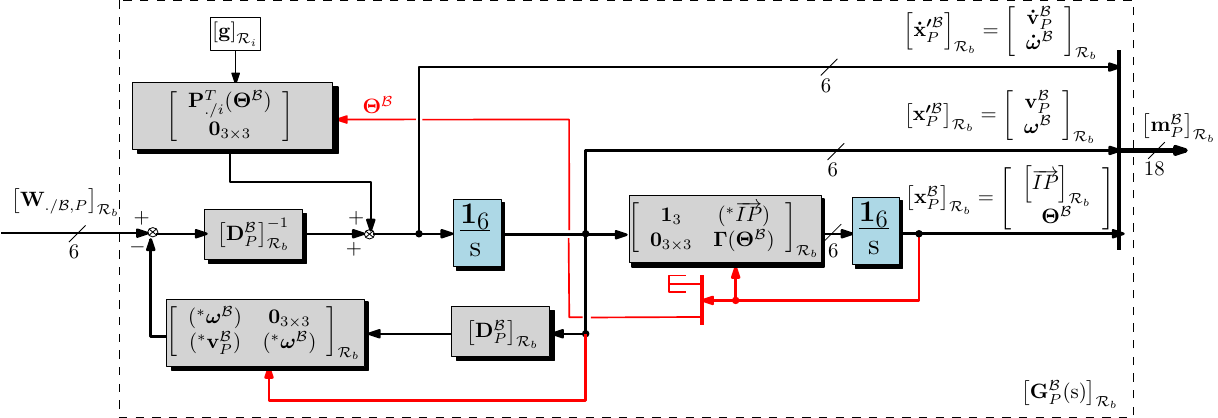}
		\caption{Block diagram representation of the inverse dynamic model $\left[\mathbf{G}^\mathcal{B}_P(\mathrm s)\right]_{\mathcal{R}_b}$ of the body $\mathcal B$ at the point $P$ and projected in $\mathcal R_b$
			.}
		\label{fig:NE_atP_inRb_18s_g_MUBO}
		\end{center}
\end{figure}

The equations of motions \eqref{eq:NE_short_g} and \eqref{eq:kin}, projected in the body frame $\mathcal R_b$ can then be represented by:
\begin{itemize}
	\item the block-diagram depicted in Figure \ref{fig:NEinv_atP_inRb_MUBO} to solve the so-called inverse dynamics (but which is based on the direct dynamic model or mass matrix  $\mathbf D^{\mathcal B}_P$). This model also highlights (in blue) the initial conditions on the twist and the pause dual vector ,
	\item the block-diagram depicted in Figure \ref{fig:NE_atP_inRb_18s_g_MUBO} to solve the forward dynamics (but which is based on the inverse dynamic model $\left[\mathbf D^{\mathcal B}_P\right]^{-1}$), denoted $\mathbf{G}^\mathcal{B}_P(\mathrm s)$. 
\end{itemize}
The model $\left[\mathbf{G}^\mathcal{B}_P(\mathrm s)\right]_{\mathcal R_b}$ is a $12$-th order model associated to the state vector $\left[{\mathbf {x'}^{\mathcal B}_P}^T\;\; {\mathbf {x}^{\mathcal B}_P}^T\right]^T_{\mathcal R_b}$. The red lines indicate the non-linear terms dependent on the state variables $\left[\mathbf {x'}^{\mathcal B}_P\right]_{\mathcal R_b}$  and $\boldsymbol \Theta^{\mathcal B}$.  For purposes of notation the projection of the pose dual vector $\mathbf {x}^{\mathcal B}_P$ and the motion vector $\mathbf m^{\mathcal B}_P$ in $\mathcal R_b$ is denoted $[.]_{\mathcal R_b}$ but the last 3 components are the components of \textsc{Euler}  angle vector $\boldsymbol \Theta^{\mathcal B}$ and do not depend on the projection frame.

In the following developments, the focus is made on the block diagram models to represent the forward dynamics but models for  inverse dynamics can also be easily derived.

\begin{figure}[!ht]
	\begin{center}
		\includegraphics[angle=0, width=12cm]{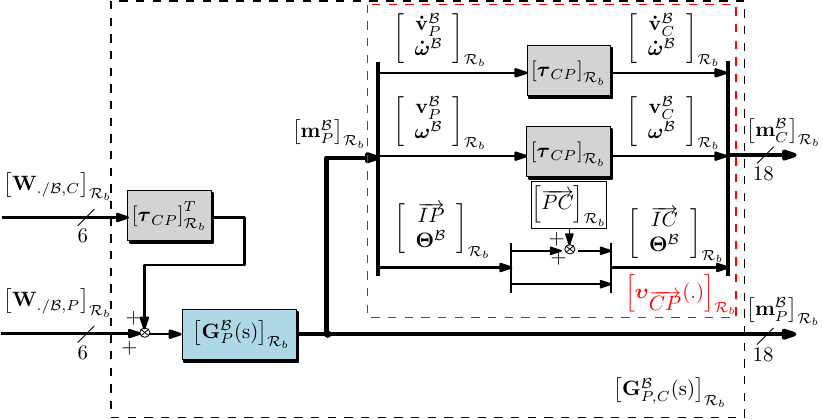}
		\caption{Block diagram representation of  $\left[\mathbf{G}^\mathcal{B}_{P,C}(\mathrm s)\right]_{\mathcal{R}_b}$: the non-linear $36\times 12$ model of a rigid body $\mathcal{B}$ at points $P$ and $C$.}
		\label{fig:NE_atPC_inRb}
	\end{center}
\end{figure}

Let us consider that  the body $\mathcal{B}$ has  $2$ connection points $P$ and $C$ where only external wrenches $\mathbf{W}_{./\mathcal{B},P}$ and $\mathbf{W}_{./\mathcal{B},C}$ can be applied, respectively. The kinematic model $\boldsymbol{\tau}_{CP}$ can be used to transport the wrench  $\mathbf{W}_{./\mathcal{B},C}$ from $C$ to $P$:
\begin{equation}
	\mathbf{W}_{ext/\mathcal B,P}= \mathbf{W}_{./\mathcal B,P}+\boldsymbol{\tau}_{CP}^T\mathbf{W}_{./\mathcal{B},C}\;.
\end{equation}
The transport of the motion vector from $P$ to $C$ can be decomposed into:
\begin{itemize}
	\item  $\boldsymbol{\mathbf x'}^{\mathcal B}_{C}=\boldsymbol{\tau}_{CP}\boldsymbol{\mathbf x'}^{\mathcal B}_{ P}$ on the twist dual vector,
	\item $\boldsymbol{\dot{\mathbf x}'}^{\mathcal B}_{ C}=\boldsymbol{\tau}_{CP}\boldsymbol{\dot{\mathbf x}'}^{\mathcal B}_{ P}$ on spatial acceleration dual vector (indeed: $\overrightarrow{CP}$ is constant in $\mathcal{R}_b$ since the body is assumed rigid), 
	\item $\overrightarrow{IC}=\overrightarrow{IP}+\overrightarrow{PC}$.
\end{itemize}
The non-linear model $\left[\mathbf{G}^\mathcal{B}_{P,C}(\mathrm s)\right]_{\mathcal{R}_b}$, projected in the body frame, of the body  $\mathcal{B}$ at points $P$ and $C$ can be described by the block-diagram of Figure \ref{fig:NE_atPC_inRb}, based on the previous model $\left[\mathbf{G}^\mathcal{B}_{P}(\mathrm s)\right]_{\mathcal{R}_b}$.

\begin{definition}{Motion vector transport from point $P$ to point $C$:}\label{def:upsilon}
	
	The operation to transport the motion vector from point $P$ to point $C$ is denoted $\boldsymbol{\upsilon_{\overrightarrow{CP}}}$:
	\begin{equation}\label{eq:upsilonCP}
		\mathbf{m}^{\mathcal B}_C=\boldsymbol \upsilon_{\overrightarrow{CP}}(\mathbf{m}^{\mathcal B}_P)\;.
	\end{equation}
	This operation, in projection in the body frame $\mathcal R_b$, is represented by the red box in Fig. \ref{fig:NE_atPC_inRb}).
\end{definition}

\section{Non-linear TITOP model}\label{sec:TITOP}
The model $\mathbf{G}^\mathcal{B}_{P,C}(\mathrm s)$ presented in the previous section is now considered as the main body $\mathcal B$ of a MBS.  $\mathcal B$ is connected at the point $C$ to a multi-body sub-system, seen as a child appendage. The $6$ rigid modes of the overall system are described by the motion of $\mathcal B$ at the point $P$. The objective is to extend the TITOP model approach, previously developed for flexible MBS in the linear case 
 to  the non-linear case for rigid MBS. The non-linear TITOP model will gather the inverse dynamics at the parent port (from the motion vector imposed  by the parent body to the applied wrench) and the forward dynamics at the child port  (from the wrench applied by the child body to the motion vector) in the same block-diagram. In the next sections, three different joints (welding joint, prismatic joint and revolute joint) are considered for the connection between  the bodies.

\subsection{Welding joint}
\begin{figure}[!ht]
	\begin{center}
		\includegraphics[angle=0, width=0.85\textwidth]{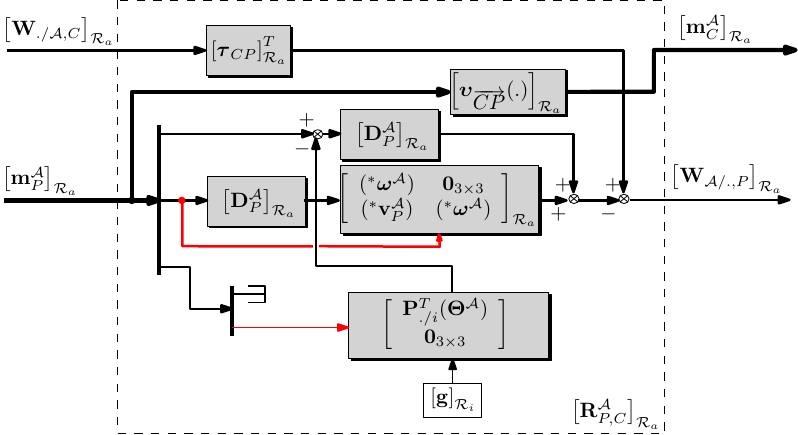}
		\caption{Block diagram model of  $\left[\mathbf{R}^\mathcal{A}_{P,C}\right]_{\mathcal{R}_a}$:  non-linear TITOP  model of a rigid body $\mathcal{A}$ at points $P$ and $C$.}
		\label{fig:NE_2port_0}
	\end{center}
\end{figure}

Let us consider a rigid appendage $\mathcal{A}$, its body frame $\mathcal{R}_a$, clamped to a parent body at point $P$ and to a child body at point $C$. The Two-Input Two-Output Port (TITOP) non-linear model $\mathbf{R}^\mathcal{A}_{P,C}$ is  the $24\times 24$ transfer between:
\begin{itemize}
	\item 2 inputs:
	\begin{itemize}
		\item the 6 component wrench $\left[\mathbf{W}_{./\mathcal{A},C}\right]_{\mathcal{R}_a}$ applied by the child body to the appendage $\mathcal{A}$ at point $C$,
		\item the 18-component motion vector $\left[\mathbf{m}^{\mathcal A}_P\right]_{\mathcal{R}_a}$ imposed by the parent body at point $P$,
	\end{itemize}
	\item 2 outputs:
	\begin{itemize}
		\item the 18-component motion vector $\left[\mathbf{m}^{\mathcal A}_C\right]_{\mathcal{R}_a}$ at point $C$,
		\item the 6 component wrench $\left[\mathbf{W}_{\mathcal{A}/.,P}\right]_{\mathcal{R}_a}$ applied by the appendage $\mathcal{A}$ to the parent body at point $P$,
	\end{itemize}
\end{itemize}
and could be represented by the block-diagram model depicted in Figure \ref{fig:NE_2port_0}. This model is directly based on the \textsc{Newton-Euler} equation \eqref{eq:NE_short_g}  applied to the body $\mathcal A$:
\begin{equation}
	\mathbf W_{./\mathcal{A},P}+\boldsymbol{\tau}_{CP}^T\mathbf{W}_{./\mathcal{A},C} =\textbf{D}_P^{\mathcal A}\left( \boldsymbol{\dot{\mathbf x}'}^{\mathcal A}_P-\left[\begin{array}{c}\mathbf g \\ \mathbf 0_{3 \times 1}\end{array}\right]\right) + \boldsymbol{\mathcal C}(\mathbf {x'}^{\mathcal A}_P)\textbf{D}_P^{\mathcal A} 
	\boldsymbol{\mathbf x'}^{\mathcal B}_P
\end{equation}
and  the motion transport operator $\boldsymbol{\upsilon_{CP}}$ defined in Definition \ref{def:upsilon}.
This model is static and does not involves additional states.

\begin{figure}[!ht]
	\begin{center}
		\includegraphics[angle=0, width=6cm]{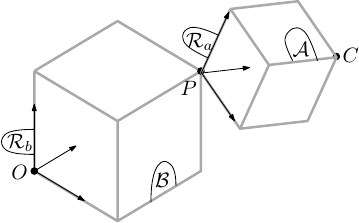}
		\caption{Two rigid bodies connected at point $P$.}
		\label{fig:2rigidbodies}
	\end{center}
\end{figure}

To build the model of the system composed of the body $\mathcal{A}$ welded to a body $\mathcal{B}$ at the point $P$ (see Figure \ref{fig:2rigidbodies}), the DCM $\mathbf{P}_{a/b}$ between the frame $\mathcal{R}_a=(P,\mathbf{x}_a,\mathbf{y}_a,\mathbf{z}_a)$ (where is described the TITOP model $\left[\mathbf{R}^\mathcal{A}_{P,C}\right]_{\mathcal{R}_a}$ of body $\mathcal{A}$) and $\mathcal{R}_b=(O,\mathbf{x}_b,\mathbf{y}_b,\mathbf{z}_b)$ (where is described the model $\left[\mathbf{G}^\mathcal{B}_{O,P}(\mathrm s)\right]_{\mathcal{R}_b}$ of body $\mathcal{B}$) must be taken into account. One can thus defines the augmented DCM: $\mathbf{P}_{a/b}^{\times 2}$ and $\mathbf P_{a/b}^{18}$  (see also nomenclature) to  transform the wrenches (or  dual  vectors) and motion vectors  from one frame to the other.  Note also that the motion vector $\mathbf m^{\mathcal A}_P$ of the body $\mathcal A$ at the point $P$ is the same as the motion vector $\mathbf m^{\mathcal B}_P$  of the body $\mathcal B$ at the same point $P$ except for the \textsc{Euler} angle vector, which must take into account the DCM $\mathbf{P}_{a/b}$: $\boldsymbol \Theta^{\mathcal A}=\boldsymbol \Theta\left(\mathbf{P}_{./i}(\boldsymbol \Theta^{\mathcal B}) \mathbf{P}_{a/b}\right)$.  

\begin{figure}[!ht]
	\begin{center}
		\includegraphics[angle=0, width=10cm]{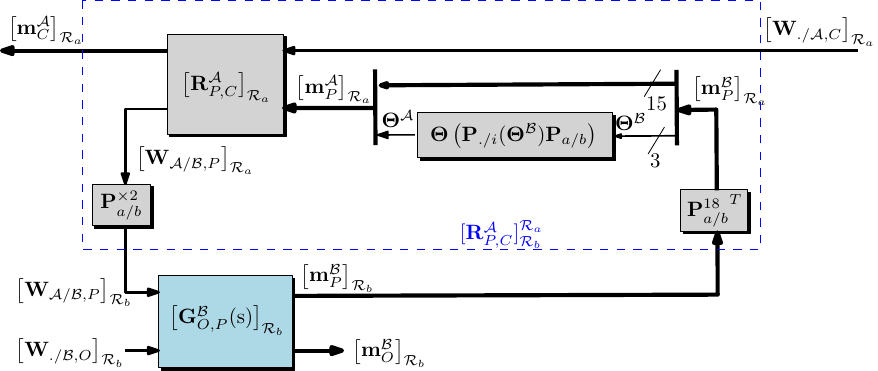}
		\caption{Block-diagram of $\left[\mathbf{G}^{\mathcal{B}+\mathcal{A}}_{O,C}(\mathrm s)\right]_{\mathcal{R}_b}^{\mathcal{R}_a}$.}
		\label{fig:2rigidbodiesG}
	\end{center}
\end{figure}

Finally, the model $\left[\mathbf{G}^{\mathcal{B}+\mathcal{A}}_{O,C}(\mathrm s)\right]_{\mathcal{R}_b}^{\mathcal{R}_a}$ of the composite $\mathcal{B}+\mathcal{A}$ at points $O$ and $C$, projected in the frame $\mathcal{R}_b$ for the lower (port $O$) channel and in frame $\mathcal{R}_a$ for the upper (port $C$) channel is described by the block-diagram depicted in Figure \ref{fig:2rigidbodiesG}.

 One can also define the model $\left[\mathbf{R}^{\mathcal{A}}_{P,C}(\mathrm s)\right]_{\mathcal{R}_b}^{\mathcal{R}_a}$ in order to include the change of frame operations as proposed in Figure  \ref{fig:2rigidbodiesG}. That will be required in the next section where the same TITOP approach is applied to a body $\mathcal A$ connected to a body $\mathcal B$ through a revolute joint. Indeed the DCM $\mathbf{P}_{a/b}$ will depend on the internal state of the revolute joint.
 
\subsection{Revolute joint}
\begin{figure}[!ht]
	\begin{center}
		\includegraphics[angle=0, width=6cm]{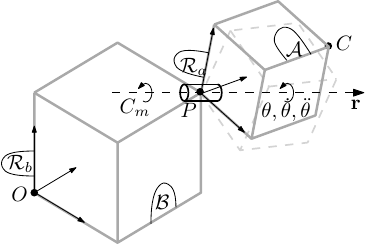}
		\caption{2 rigid bodies connected at point $P$ through a revolute joint around axis $\mathbf{r}$.}
		\label{fig:2rigidbodiesrevolute}
	\end{center}
\end{figure}

Let $\theta$, $\dot{\theta}$, $\ddot\theta$ the angular configuration, rate and acceleration inside the revolute joint between bodies $\mathcal{B}$ and $\mathcal{A}$ at the connection point $P$ (see Figure \ref{fig:2rigidbodiesrevolute}). In the following developments the revolute joint belongs to the body $\mathcal{A}$. Thus the joint axis $\mathbf r$ is expressed in $\mathcal{R}_a$: $[\mathbf r]_{\mathcal{R}_a}$. $C_m$ is the driving torque applied on body $\mathcal{A}$ inside the revolute joint by a driving mechanism.

\begin{figure}[!ht]
	\begin{center}
		\includegraphics[angle=0, width=0.6\textwidth]{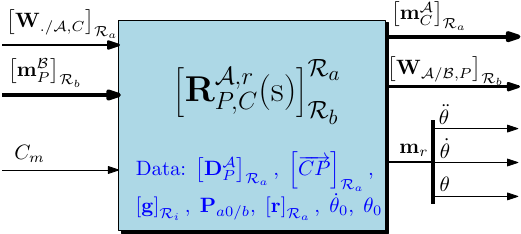}
		\caption{Block-diagram representation of $[\mathbf{R}^{\mathcal{A},r}_{P,C}(\mathrm s)]^{\mathcal{R}_a}_{\mathcal{R}_b}$: non-linear TITOP model of $\mathcal{A}$ augmented with the transfer from the joint driving torque $C_m$ to the joint relative motion $\mathbf m_r$.}
		\label{fig:NE_2port_1}
	\end{center}
\end{figure}

The objective is to compute the model $\left[\mathbf{R}^{\mathcal A,r}_{P,C}(\mathrm s )\right]_{\mathcal{R}_b}^{\mathcal R_a}$ of the body $\mathcal A$ with its revolute joint as defined in Figure \ref{fig:NE_2port_1}. In addition to the inverse channel at the port $C$ (i.e.: the transfer from $\left[\mathbf{W}_{./\mathcal{A},C}\right]_{\mathcal{R}_a}$ to $\left[\mathbf{m}^{\mathcal A}_C\right]_{\mathcal{R}_a}$) and to the direct channel at the port $P$ (i.e.: the transfer from $\left[\mathbf{m}^{\mathcal B}_P\right]_{\mathcal{R}_b}$ to $\left[\mathbf{W}_{\mathcal{A/B},P}\right]_{\mathcal{R}_b}$), as already defined for the welding joint in Figure \ref{fig:2rigidbodiesG}, this model also includes the channel from the torque $C_m$ to the revolute joint relative motion $\mathbf m_r=\left[\ddot\theta,\;\dot\theta,\;\theta\right]^T$. This $27\times 25$ model is a second order model which depends on the following parameters:
\begin{itemize}
	\item $\left[\mathbf{D}^{\mathcal A}_{P}\right]_{\mathcal{R}_a}$, $\left[\overrightarrow{CP}\right]_{\mathcal R_a}$, $\left[\mathbf r\right]_{\mathcal R_a}$: the dynamic model of the body $\mathcal A$ at the point $P$, the position of the point $C$  w.r.t. $P$ and the direction of the revolute joint. All these data are projected in $\mathcal R_a$,
	\item $\left[\mathbf g\right]_{\mathcal R_i}$: the uniform gravitational acceleration in the inertial frame,
	\item $\mathbf P_{a0/b}$: the DCM from $\mathcal R_a$ to $\mathcal R_b$ when the angular configuration of the revolute joint is null ($\theta=0$),
	\item $\theta_0$ and $\dot\theta_0$:  the initial angular position and velocity of the revolute joint (i.e.: the initial state of this second order model).
\end{itemize}

First of all, the angular configuration $\theta$ around $\mathbf{r}$ must be taken into account in the DCM from $\mathcal R_a$ to $\mathcal R_b$, now denoted $\mathbf{P}_{a\theta/b}$, and the augmented  DCM $\mathbf{P}_{a\theta/b}^{\times 2}$ and $\mathbf{P}_{a\theta/b}^{18}$: 
\begin{equation}\label{eq:Pa/b_rev}
	\mathbf{P}_{a\theta/b}= \mathbf{P}_{a0/b}\,\mathrm{e}^{\theta(^*[\widetilde{\mathbf r}])_{\mathcal{R}_a}}
\end{equation}
where $\widetilde{\mathbf r}$ is a unit vector along $\mathbf{r}$.

Then, one can express the motion vector $\mathbf m^{\mathcal A}_{P}$ at point $P$ from the motion vector $\mathbf m^{\mathcal B}_P$ at $P$ and $\theta$, $\dot{\theta}$, $\ddot \theta$. Indeed:
\[
\boldsymbol \Theta^{\mathcal A}=\boldsymbol \Theta\left(\mathbf{P}_{./i}(\boldsymbol \Theta^{\mathcal B}) \mathbf{P}_{a\theta/b}\right)
\]
\[
\mathbf v^{\mathcal A}_{P} = \mathbf v^{\mathcal B}_{P}
\]
\[
\boldsymbol{\omega}^{\mathcal A}=\boldsymbol{\omega}^{\mathcal B}+\dot{\theta}\widetilde{\mathbf r}
\]
\[
\dot{\mathbf v}^{\mathcal A}_{P} = \left.\frac{d \mathbf v^{\mathcal B}_{P}}{dt}\right\vert_{\mathcal{R}_a} = \dot{\mathbf v}^{\mathcal B}_{P} -\dot{\theta}(^*\widetilde{\mathbf r})\mathbf v^{\mathcal B}_{P}
\]
\[
\boldsymbol{\dot\omega}^{\mathcal A}=\left.\frac{d \boldsymbol{\omega}^{\mathcal A} }{dt}\right\vert_{\mathcal{R}_a} = \left.\frac{d \boldsymbol{\omega}^{\mathcal B} }{dt}\right\vert_{\mathcal{R}_a}+\ddot{\theta} \widetilde{\mathbf r}=\boldsymbol{\dot\omega}^{\mathcal B}-\dot{\theta}(^*\widetilde{\mathbf r})\boldsymbol{\omega}^{\mathcal B}+\ddot{\theta} \widetilde{\mathbf r}
\]
\begin{equation}\mbox{Thus: }\label{eq:transportrev}
	\mathbf m^{\mathcal A}_{P}
	=
	\left[\begin{array}{l}
		\boldsymbol{\dot{\mathbf x}'}^{\mathcal A}_{P}=\boldsymbol{\dot{\mathbf x}'}^{\mathcal B}_{P}+\left[\begin{array}{l} -\dot{\theta}(^*\widetilde{\mathbf r})\mathbf v^{\mathcal B}_{P}\\ \ddot{\theta} \widetilde{\mathbf r} -\dot{\theta}(^*\widetilde{\mathbf r})\boldsymbol{\omega}^{\mathcal B}\end{array}\right] \\
		
		\boldsymbol{\mathbf x'}^{\mathcal A}_{P}=\boldsymbol{\mathbf x'}^{\mathcal B}_{P}+\left[\begin{array}{l} \mathbf 0_{3 \times 1}\\ \dot{\theta} \widetilde{\mathbf r}\end{array}\right] \\
		
		\boldsymbol{\mathbf x}^{\mathcal A}_{P}=\left[\begin{array}{l} \overrightarrow{IP} \\ \boldsymbol \Theta^{\mathcal A}=\boldsymbol \Theta^{\mathcal A}\left(\mathbf{P}_{./i}(\boldsymbol \Theta^{\mathcal B}) \mathbf{P}_{a\theta/b}\right)\end{array}\right]
		
	\end{array}\right]\;.
\end{equation}

The \textsc{Newton-Euler} model  \eqref{eq:NE_short_g}, applied to the  appendage $\mathcal A$ loaded by external wrenches at point $P$ and $C$, reads:
\begin{eqnarray}\nonumber
	& &- \mathbf W_{\mathcal{A/B},P}+\boldsymbol \tau_{CP}^T\mathbf W_{./\mathcal{A},C}=\mathbf{D}_{P}^{\mathcal A}
	\left(\boldsymbol{\dot{\mathbf x}'}^{\mathcal A}_{P} - \left[\begin{array}{l} \mathbf g \\ \mathbf 0_{3 \times 1}\end{array}\right] \right)+ 
	\boldsymbol{\mathcal{C}}(\mathbf{x'}^{\mathcal A}_{P})
	\mathbf{D}_{P}^{\mathcal A} 
	\boldsymbol{\mathbf x'}^{\mathcal A}_{P}\\ \label{eq:dynrev}
	&=&\mathbf{D}_{P}^{\mathcal A}\left(
	\boldsymbol{\dot{\mathbf x}'}^{\mathcal B}_{P}- \left[\begin{array}{l} \mathbf g \\ \mathbf 0_{3 \times 1}\end{array}\right] +\ddot{\theta}\left[\begin{array}{l}  \mathbf 0_{3 \times 1} \\ \widetilde{\mathbf r}\end{array}\right] -\dot{\theta}\left[\begin{array}{c}(^*\widetilde{\mathbf r})\textbf v^{\mathcal B}_P\\ (^*\widetilde{\mathbf r})\boldsymbol{\omega}^{\mathcal B}
	\end{array} \right]\right) + 
	\boldsymbol{\mathcal{C}}(\mathbf{x'}^{\mathcal A}_{P})
	\mathbf{D}_{P}^{\mathcal A} 
	\boldsymbol{\mathbf x'}^{\mathcal A}_{P}
\end{eqnarray}

The driving torque $C_m$ is the projection  along $\mathbf r$ of the torque $\mathbf T_{\mathcal{B/A},P}$ applied by $\mathcal B$ on $\mathcal{A}$ at $P$:
\begin{equation}\label{eq:Cm}
	C_m=\left[\begin{array}{c}
		\mathbf{0}_{3 \times 1} \\ \widetilde{\mathbf r}
	\end{array}\right]^T\mathbf W_{\mathcal{B/A},P}=-\left[\begin{array}{c}
	\mathbf{0}_{3 \times 1} \\ \widetilde{\mathbf r}
	\end{array}\right]^T\mathbf W_{\mathcal{A/B},P}\;.
\end{equation}

Then pre-multiplying \eqref{eq:dynrev} by $\left[\begin{array}{c}
	\mathbf{0}_{3 \times 1} \\ \widetilde{\mathbf r}
\end{array}\right]^T$ and denoting :
\begin{equation}\label{eq:Jr}
	J_r=\left[\begin{array}{c}
		\mathbf{0}_{3 \times 1} \\ \widetilde{\mathbf r}
	\end{array}\right]^T \mathbf{D}_{P}^{\mathcal A}  \left[\begin{array}{c}
	\mathbf{0}_{3 \times 1} \\ \widetilde{\mathbf r}
	\end{array}\right]
\end{equation}
the apparent inertia of the body $\mathcal A$ seen from the revolute joint, one can express:
\begin{eqnarray}\nonumber
C_m=J_r\ddot\theta+ \left[\begin{array}{c}
	\mathbf{0}_{3 \times 1} \\ \widetilde{\mathbf r}
\end{array}\right]^T& &\left(\mathbf{D}_{P}^{\mathcal A}\left(
\boldsymbol{\dot{\mathbf x}'}^{\mathcal B}_{P}- \left[\begin{array}{l} \mathbf g \\ \mathbf 0_{3 \times 1}\end{array}\right] -\dot{\theta}\left[\begin{array}{c}(^*\widetilde{\mathbf r})\textbf v^{\mathcal B}_P\\ (^*\widetilde{\mathbf r})\boldsymbol{\omega}^{\mathcal B}
\end{array} \right]\right)\right.  \\ 
& &\;\;\left. + \boldsymbol{\mathcal{C}}(\mathbf{x'}^{\mathcal A}_{P})
\mathbf{D}_{P}^{\mathcal A} 
\boldsymbol{\mathbf x'}^{\mathcal A}_{P}-\boldsymbol \tau_{CP}^T\mathbf W_{./\mathcal{A},C}\right)\;.
\end{eqnarray}
Thus, the in-joint acceleration $\ddot\theta$ to be integrated twice from the initial conditions $\dot\theta_0$ and $\theta_0$ is:
\begin{eqnarray}\nonumber
	\ddot\theta=\frac{C_m}{J_r}- \frac{1}{J_r}\left[\begin{array}{c}
		\mathbf{0}_{3 \times 1} \\ \widetilde{\mathbf r}
	\end{array}\right]^T& &\left(\mathbf{D}_{P}^{\mathcal A}\left(
	\boldsymbol{\dot{\mathbf x}'}^{\mathcal B}_{P}- \left[\begin{array}{l} \mathbf g \\ \mathbf 0_{3 \times 1}\end{array}\right] -\dot{\theta}\left[\begin{array}{c}(^*\widetilde{\mathbf r})\textbf v^{\mathcal B}_P\\ (^*\widetilde{\mathbf r})\boldsymbol{\omega}^{\mathcal B}
	\end{array} \right]\right)\right.  \\  \label{eq:thetaddot}
	& &\;\;\left. + \boldsymbol{\mathcal{C}}(\mathbf{x'}^{\mathcal A}_{P})
	\mathbf{D}_{P}^{\mathcal A} 
	\boldsymbol{\mathbf x'}^{\mathcal A}_{P}-\boldsymbol \tau_{CP}^T\mathbf W_{./\mathcal{A},C}\right)\;.
\end{eqnarray}

\begin{figure}[!ht]
	\begin{center}
		\includegraphics[angle=0, width=\textwidth]{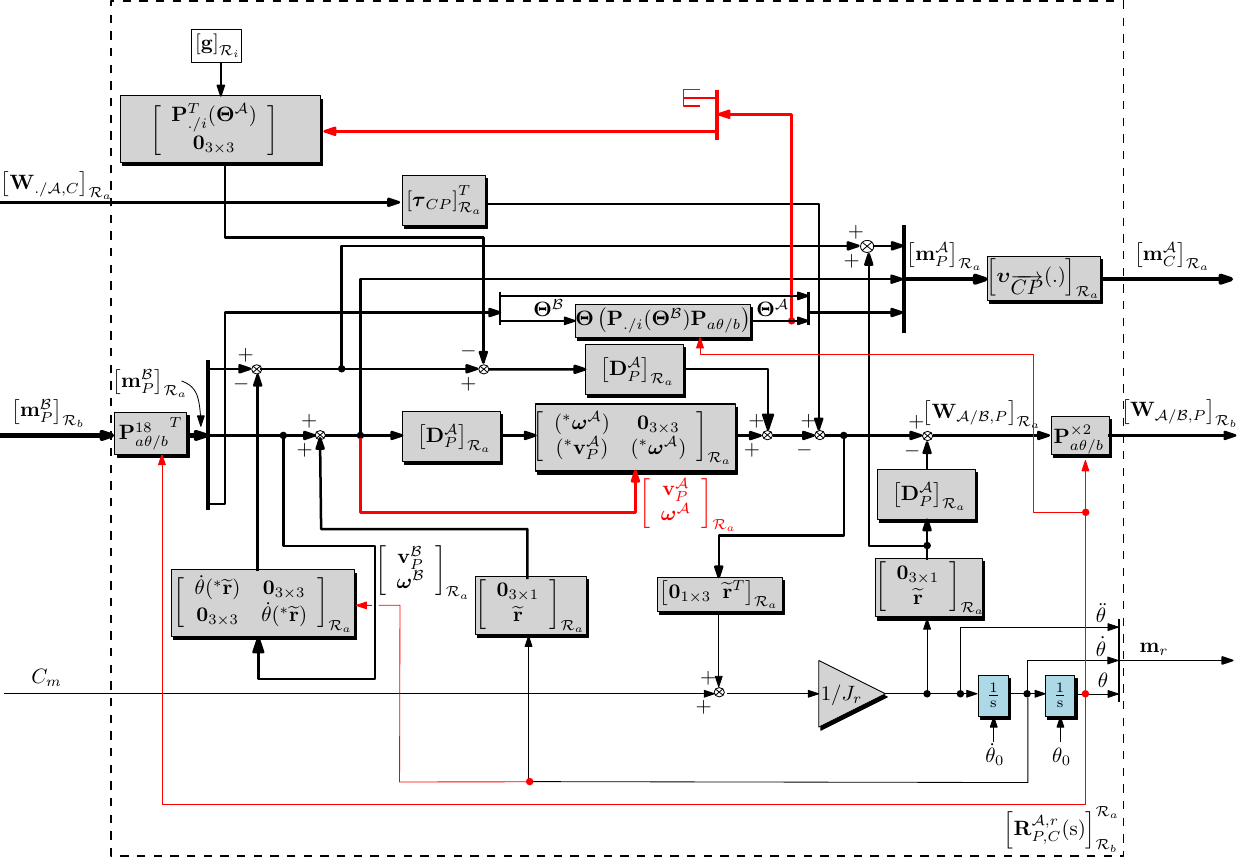}
		\caption{Detailed block-diagram model of $[\mathbf{R}^{\mathcal{A},r}_{P,C}(\mathrm s)]^{\mathcal{R}_a}_{\mathcal{R}_b}$.}
		\label{fig:NE_2port_rev_2_g_d}
	\end{center}
\end{figure}

Finally, the motion vector $\mathbf m^{\mathcal A}_C$ of the body $\mathcal A$ at the point $C$ can be transported from the point $P$ (see definition \eqref{def:upsilon}):
	\begin{equation}\label{eq:upsilonCP_A}
	\mathbf{m}^{\mathcal A}_C=\boldsymbol\upsilon_{\overrightarrow{CP}}(\mathbf{m}^{\mathcal A}_P)\;.
\end{equation}
From the DCM $\mathbf P_{a\theta/b}$ defined in \eqref{eq:Pa/b_rev},  $\mathbf P_{a\theta/b}^{\times 2}$,  $\mathbf P_{a\theta/b}^{18}$ and the inertia $J_r$ defined in \eqref{eq:Jr}, the model $[\mathbf{R}^{\mathcal{A},r}_{P,C}(\mathrm s)]^{\mathcal{R}_a}_{\mathcal{R}_b}$ is entirely defined by equations \eqref{eq:transportrev}, \eqref{eq:dynrev}, \eqref{eq:thetaddot} and \eqref{eq:upsilonCP_A} which must be projected in the body frame $\mathcal R_a$ and can be represented by the block-diagram model depicted in Figure \ref{fig:NE_2port_rev_2_g_d}.

\subsection{Prismatic joint}
\begin{figure}[!ht]
	\begin{center}
		\includegraphics[angle=0, width=6cm]{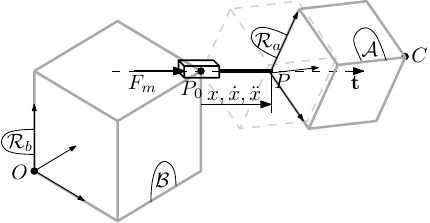}
		\caption{2 rigid bodies connected at point $P_0$ through a prismatic joint along axis $\mathbf{t}$.}
		\label{fig:2rigidbodiesprismatic}
	\end{center}
\end{figure}

Let $x$, $\dot{x}$, $\ddot x$ the linear configuration, velocity and acceleration inside the prismatic joint between bodies $\mathcal{B}$ and $\mathcal{A}$ at the connection point $P_0$ (see Figure \ref{fig:2rigidbodiesprismatic}). In the following developments the prismatic joint belongs to the body $\mathcal{A}$. Thus the joint axis $\mathbf t$ is expressed in the $\mathcal{R}_a$: $[\mathbf t]_{\mathcal{R}_a}$. The points $P_0$ and $P$ are fixed in frames $\mathcal R_b$ and $\mathcal R_a$, respectively. $P\equiv P_0$ in the nominal configuration $x=0$. $F_m$ is the driving force applied on body $\mathcal{A}$ inside the prismatic joint by a driving mechanism.

\begin{figure}[!ht]
	\begin{center}
		\includegraphics[angle=0, width=\textwidth]{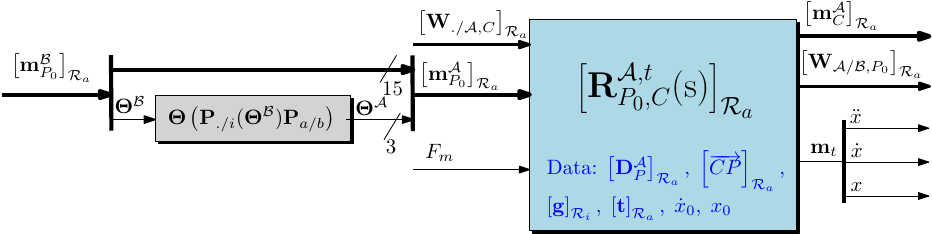}
		\caption{Block-diagram representation of $[\mathbf{R}^{\mathcal{A},t}_{P_0,C}(\mathrm s)]_{\mathcal{R}_a}$: non-linear TITOP model of $\mathcal{A}$ augmented with the transfer from the joint driving force $F_m$ to the joint relative motion $\mathbf m_t$.}
		\label{fig:NE_2port_2}
	\end{center}
\end{figure}

Similarly to the previous case, the objective is to compute the model $\left[\mathbf{R}^{\mathcal A,t}_{P_0,C}(\mathrm s )\right]_{\mathcal{R}_a}$ of the body $\mathcal A$ with its prismatic joint as defined in Figure \ref{fig:NE_2port_2}. Note that in the prismatic joint case both inverse channel at port $C$ and the direct channel at port $P_0$ are projected in the body frame $\mathcal R_a$. The DCM $\mathbf P_{a/b}$ between the body $\mathcal A$ and the parent body $\mathcal B$ can be taken into account exactly as depicted in Figure \ref{fig:2rigidbodiesG}. 
 This $27\times 25$ model is a second ordre model which depends on the following parameters: $\left[\mathbf{D}^{\mathcal A}_{P}\right]_{\mathcal{R}_a}$, $\left[\overrightarrow{CP}\right]_{\mathcal R_a}$, $\left[\mathbf t\right]_{\mathcal R_a}$, $\left[\mathbf g\right]_{\mathcal R_i}$ and  finally $x_0$ and $\dot x_0$:  the initial linear position and velocity of the prismatic joint (i.e.: the initial state of this second order model).

First of all, one can express the motion vector $\mathbf m^{\mathcal A}_{P}$ at point $P$ from the motion vector $\mathbf m^{\mathcal A}_{P_0}$ at $P_0$ and $x$, $\dot{x}$, $\ddot x$. Indeed:
\[
\overrightarrow{IP}  = \overrightarrow{IP_0} + x\widetilde{\mathbf t}
\]
\[
\mathbf v^{\mathcal A}_{P} = \left.\frac{d\overrightarrow{IP}}{dt}\right\vert_{\mathcal{R}_i}  =  \mathbf v^{\mathcal A}_{P_0} -x(^*\widetilde{\mathbf t})\boldsymbol{\omega}^{\mathcal A}+\dot{x}\widetilde{\mathbf t}
\]
\[
\dot{\mathbf v}^{\mathcal A}_{P} = \left.\frac{d \mathbf v^{\mathcal A}_{P}}{dt}\right\vert_{\mathcal{R}_a} =\dot{\mathbf v}^{\mathcal A}_{P_0}- x(^*\widetilde{\mathbf t}) \boldsymbol{\dot{\omega}}^{\mathcal A}-\dot{x}(^*\widetilde{\mathbf t})\boldsymbol{\omega}^{\mathcal A}+\ddot{x}\widetilde{\mathbf t}
\]
where $\widetilde{\mathbf t}$ is a unit vector along $\mathbf{t}$. 

Thus using the operator $\boldsymbol{\upsilon}_{\overrightarrow{PP_0}}=\boldsymbol{\upsilon}_{-x\widetilde{\mathbf t}}$ defined in definition  \ref{def:upsilon}, one can write
\begin{equation}\label{eq:transportprisme}
	\mathbf m^{\mathcal A}_{P}=\boldsymbol{\upsilon}_{-x\widetilde{\mathbf t}}(\mathbf m^{\mathcal A}_{P_0})+
	\left[\begin{array}{c}
		\ddot{x}\widetilde{\mathbf t}-\dot{x}(^*\widetilde{\mathbf t})\boldsymbol{\omega}^{\mathcal A}\\
		\mathbf 0_{3\times 1}\\
		\dot{x}\widetilde{\mathbf t} \\
		\mathbf 0_{3\times 1}\\ 
		\mathbf 0_{3\times 1}\\  \mathbf 0_{3\times 1}
	\end{array}\right]
		\Rightarrow
	\left[\begin{array}{l}
		\boldsymbol{\dot{\mathbf x}'}^{\mathcal A}_{P}=\boldsymbol{\tau}_{-x\widetilde{\mathbf t}}\boldsymbol{\dot{\mathbf x}'}^{\mathcal A}_{P_0}+\left[\begin{array}{l} \ddot{x}\widetilde{\mathbf t}-\dot{x}(^*\widetilde{\mathbf t})\boldsymbol{\omega}^{\mathcal A}\\ \mathbf 0_{3 \times 1}\end{array}\right] \\
		
		\boldsymbol{\mathbf x'}^{\mathcal A}_{P}=\boldsymbol{\tau}_{-x\widetilde{\mathbf t}}\boldsymbol{\mathbf x'}^{\mathcal A}_{P_0}+\left[\begin{array}{l} \dot{x}\widetilde{\mathbf t}\\  \mathbf 0_{3 \times 1}\end{array}\right] \\
		
		\boldsymbol{\mathbf x}^{\mathcal A}_{P}=\boldsymbol{\mathbf x}^{\mathcal A}_{P_0}+\left[\begin{array}{l} x\widetilde{\mathbf t} \\ \mathbf 0_{3\times 1}\end{array}\right]
		
	\end{array}\right]\;.
\end{equation}
The \textsc{Newton-Euler} model  \eqref{eq:NE_short_g}, applied to the  appendage $\mathcal A$ loaded by external wrenches at point $P$ and $C$, reads:
\begin{eqnarray} \label{eq:dynprism}
	& &- \mathbf W_{\mathcal{A/B},P}+\boldsymbol \tau_{CP}^T\mathbf W_{./\mathcal{A},C}=\mathbf{D}_{P}^{\mathcal A}
	\left(\boldsymbol{\dot{\mathbf x}'}^{\mathcal A}_{P} - \left[\begin{array}{l} \mathbf g \\ \mathbf 0_{3 \times 1}\end{array}\right] \right)+ 
	\boldsymbol{\mathcal{C}}(\mathbf{x'}^{\mathcal A}_{P})
	\mathbf{D}_{P}^{\mathcal A} 
	\boldsymbol{\mathbf x'}^{\mathcal A}_{P}\\ \nonumber
	&=&\mathbf{D}_{P}^{\mathcal A}\left(\boldsymbol{\tau}_{-x\widetilde{\mathbf t}}
	\boldsymbol{\dot{\mathbf x}'}^{\mathcal A}_{P_0}- \left[\begin{array}{l} \mathbf g \\ \mathbf 0_{3 \times 1}\end{array}\right] +\ddot{x}\left[\begin{array}{l} \widetilde{\mathbf t} \\  \mathbf 0_{3 \times 1} \end{array}\right] -\dot{x}\left[\begin{array}{c}(^*\widetilde{\mathbf t}) \boldsymbol{\omega}^{\mathcal A}\\  \mathbf 0_{3 \times 1}
	\end{array} \right]\right) + 
	\boldsymbol{\mathcal{C}}(\mathbf{x'}^{\mathcal A}_{P})
	\mathbf{D}_{P}^{\mathcal A} 
	\boldsymbol{\mathbf x'}^{\mathcal A}_{P}
\end{eqnarray}

The driving force $F_m$ is the projection  along $\mathbf t$ of the force $\mathbf F_{\mathcal{B/A},P}$ applied by $\mathcal B$ on $\mathcal{A}$ at $P$:
\begin{equation}\label{eq:Cm}
	F_m=\left[\begin{array}{c}
	 \widetilde{\mathbf t} \\ \mathbf{0}_{3 \times 1}
	\end{array}\right]^T\mathbf W_{\mathcal{B/A},P}=-\left[\begin{array}{c}
	\widetilde{\mathbf t} \\ \mathbf{0}_{3 \times 1}
	\end{array}\right]^T\mathbf W_{\mathcal{A/B},P}\;.
\end{equation}
Then pre-multiplying \eqref{eq:dynprism} by $\left[\begin{array}{c}
\widetilde{\mathbf t} \\	\mathbf{0}_{3 \times 1} 
\end{array}\right]^T$ and denoting $m^{\mathcal A}$ the mass of the body $\mathcal A$, one can express:
\begin{eqnarray}\nonumber
	F_m=m^{\mathcal A}\ddot x+ \left[\begin{array}{c}
	\widetilde{\mathbf t} \\	\mathbf{0}_{3 \times 1}
	\end{array}\right]^T& &\left(\mathbf{D}_{P}^{\mathcal A}
	\left(\boldsymbol{\tau}_{-x\widetilde{\mathbf t}}
	\boldsymbol{\dot{\mathbf x}'}^{\mathcal A}_{P_0}- \left[\begin{array}{l} \mathbf g \\ \mathbf 0_{3 \times 1}\end{array}\right]  -\dot{x}\left[\begin{array}{c}(^*\widetilde{\mathbf t}) \boldsymbol{\omega}^{\mathcal A}\\  \mathbf 0_{3 \times 1}
	\end{array} \right]\right)
	\right.  \\ 
	& &\;\;\left. + \boldsymbol{\mathcal{C}}(\mathbf{x'}^{\mathcal A}_{P})
	\mathbf{D}_{P}^{\mathcal A} 
	\boldsymbol{\mathbf x'}^{\mathcal A}_{P}-\boldsymbol \tau_{CP}^T\mathbf W_{./\mathcal{A},C}\right)\;.
\end{eqnarray}
Thus, the in-joint acceleration $\ddot x$ to be integrated twice from the initial conditions $\dot x_0$ and $x_0$ is:
\begin{eqnarray}\nonumber
	\ddot x=\frac{F_m}{m^{\mathcal A}}- \frac{1}{m^{\mathcal A}}\left[\begin{array}{c}
		\widetilde{\mathbf t} \\	\mathbf{0}_{3 \times 1}
	\end{array}\right]^T& &\left(\mathbf{D}_{P}^{\mathcal A}
		\left(\boldsymbol{\tau}_{-x\widetilde{\mathbf t}}
	\boldsymbol{\dot{\mathbf x}'}^{\mathcal A}_{P_0}- \left[\begin{array}{l} \mathbf g \\ \mathbf 0_{3 \times 1}\end{array}\right]  -\dot{x}\left[\begin{array}{c}(^*\widetilde{\mathbf t}) \boldsymbol{\omega}^{\mathcal A}\\  \mathbf 0_{3 \times 1}
	\end{array} \right]\right)
	\right.  \\  \label{eq:xddot}
	& &\;\;\left. + \boldsymbol{\mathcal{C}}(\mathbf{x'}^{\mathcal A}_{P})
	\mathbf{D}_{P}^{\mathcal A} 
	\boldsymbol{\mathbf x'}^{\mathcal A}_{P}-\boldsymbol \tau_{CP}^T\mathbf W_{./\mathcal{A},C}\right)\;.
\end{eqnarray}
Finally, the wrench applied by $\mathcal{A}$ on $\mathcal B$ at point $P_0$ is:
\begin{equation}\label{eq:wA/.,P_0}
	\mathbf W_{\mathcal{A/B},P_0}=\boldsymbol \tau_{-x\widetilde{\mathbf t}}^T\mathbf W_{\mathcal{A/B},P}\;.
\end{equation}

\begin{figure}[!ht]
	\begin{center}
	\includegraphics[angle=0, width=\textwidth]{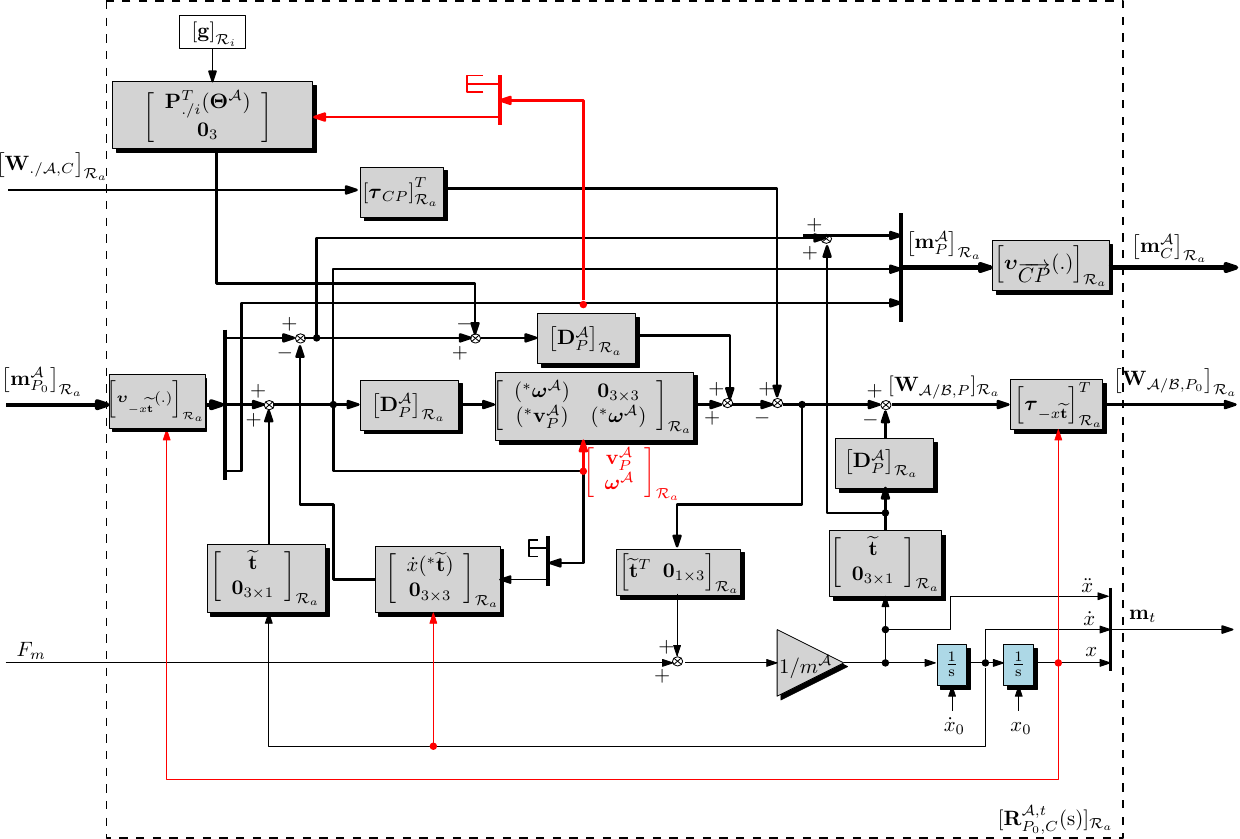}
\caption{Block-diagram representation of $[\mathbf{R}^{\mathcal{A},t}_{P_0,C}(\mathrm s)]_{\mathcal{R}_a}$: non-linear TITOP model of $\mathcal{A}$ augmented with the transfer from the joint driving force $F_m$ and the joint relative motion $\mathbf m_t$.}
\label{fig:NE_2port_pris_m1}
	\end{center}
\end{figure}

The model $[\mathbf{R}^{\mathcal{A},t}_{P_0,C}(\mathrm s)]_{\mathcal{R}_a}$ is entirely defined by equations \eqref{eq:transportprisme}, \eqref{eq:dynprism}, \eqref{eq:xddot}, \eqref{eq:upsilonCP_A} and \eqref{eq:wA/.,P_0} which must be projected in the body frame $\mathcal R_a$ and can be represented by the block-diagram model depicted in Figure \ref{fig:NE_2port_pris_m1}.

Note that these models, respectively in the revolute joint  and the prismatic joint cases, can be easily extended to the models $[\mathbf{R}^{\mathcal{A},r}_{P,C_1,\cdots C_n}(\mathrm s)]^{\mathcal{R}_a}_{\mathcal{R}_b}$  and  $[\mathbf{R}^{\mathcal{A},t}_{P_0,C_1,\cdots C_n}(\mathrm s)]_{\mathcal{R}_a}$ if $n$ child bodies $\mathcal C_i,\;i=1,\cdots n$ are connected to $\mathcal A$ at the point $C_i$.  Indeed the geometric models $\boldsymbol{\tau}_{C_iP}$,  to be applied on the wrenches, and the transport operations $\boldsymbol\upsilon_{C_iP}$, to be applied on the motion vectors, will allow to build such multi-port models which are required to model open-kinematic trees of rigid bodies. 

\subsection{Loop closure constraints}
\begin{figure}[!ht]
	\begin{center}
		\includegraphics[angle=0, width=0.6\textwidth]{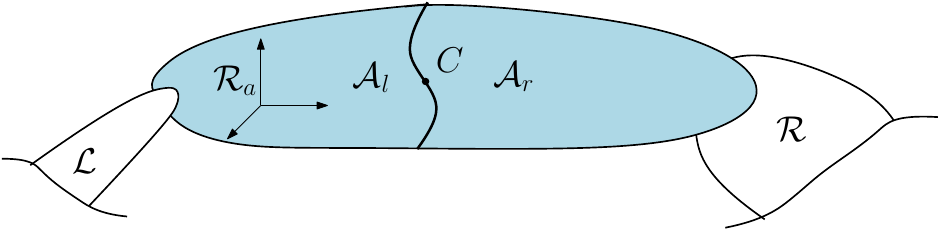}
		\caption{Opening the loop on the body $\mathcal A$ between the bodies  $\mathcal L$ and $\mathcal R$ in a multi-body system with closed kinematic chains.}
		\label{fig:Loop}
	\end{center}
\end{figure}

\begin{figure}[!ht]
	\begin{center}
		\includegraphics[angle=0, width=0.6\textwidth]{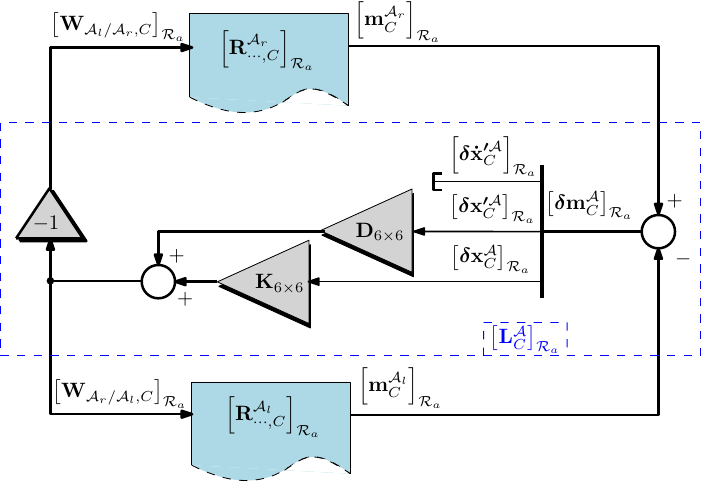}
		\caption{Definition of the loop closure block $\left[\mathbf{L}^{\mathcal A}_{C}\right]_{\mathcal{R}_a}$.}
		\label{fig:LoopBlock}
	\end{center}
\end{figure}

To model multi-body systems with closed-kinematic chains, one can break the loop at the level of a particular body $\mathcal A$ as depicted in Figure \ref{fig:Loop}. This body, inserted between the bodies $\mathcal L$ and $\mathcal R$, is split into two parts $\mathcal A_l$ and $\mathcal A_r$ at the level of the point $C$. The part $\mathcal A_l$ (resp. $\mathcal A_r$) will end the model of the $\mathcal L$ (resp $\mathcal R$) opened kinematic chain using a TITOP model $\mathbf R^{\mathcal A_l}_{\cdots,C}$ (resp. $\mathbf R^{\mathcal A_r}_{\cdots,C}$).  The upper channel of this TITOP model is the inverse transfer from the interaction wrench $\mathbf{W}_{\mathcal{A}_r/\mathcal{A}_l,C}$ (resp. $\mathbf{W}_{\mathcal{A}_l/\mathcal{A}_r,C}$) to the motion vector $\mathbf{m}^{\mathcal A_l}_C$ (resp. $\mathbf{m}^{\mathcal A_r}_C$) of the point $C$ projected in the body frame $\mathcal R_a$. One  can then close the loop assuming the body $\mathcal A$ have a mechanical impedance modeled as a local spring-damper at the point $C$. As depicted in Figure \ref{fig:LoopBlock}, this model, named $\mathbf{L}^{\mathcal A}_{C}$, feedbacks the relative motion vector $\boldsymbol{\delta \mathbf m}^{\mathcal A}_{C}=\boldsymbol{\mathbf m}^{\mathcal A_r}_{C}-\boldsymbol{\mathbf m}^{\mathcal A_l}_{C}$  at the point $C$ to the interaction wrench through the $6 \times 6$ stiffness matrix $\mathbf K$ and the $6 \times 6$ damping matrix $\mathbf D$.

\section{Illustration and validation}
\begin{figure}[!ht]
	\begin{center}
		\includegraphics[angle=0, width=0.7\textwidth]{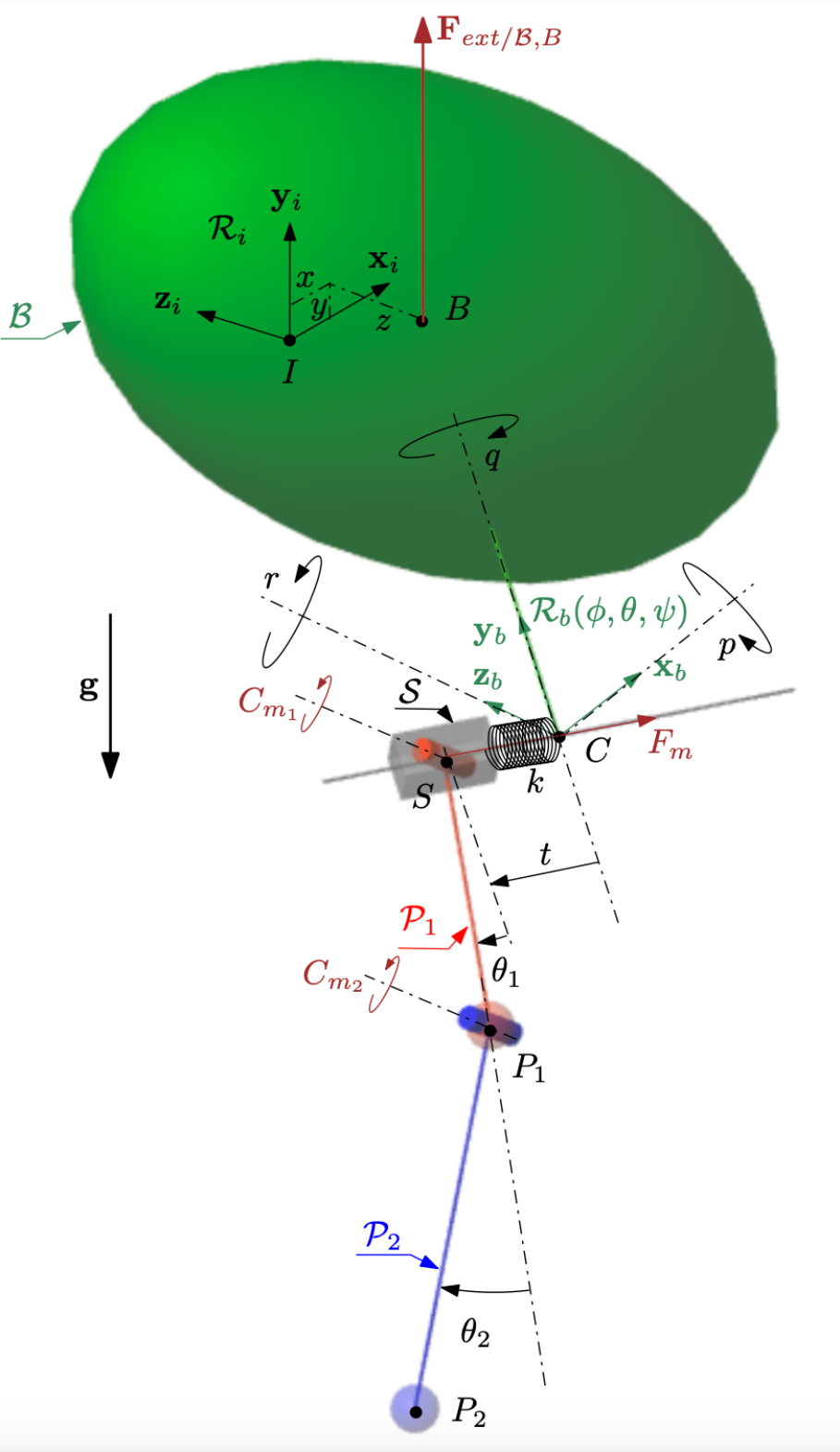}
		\caption{The Mechanical Explorer model of the system $\mathcal G$.}
		\label{fig:schema1}
	\end{center}
\end{figure}

\begin{figure}[!ht]
	\begin{center}
			\includegraphics[angle=0, width=0.9\textwidth]{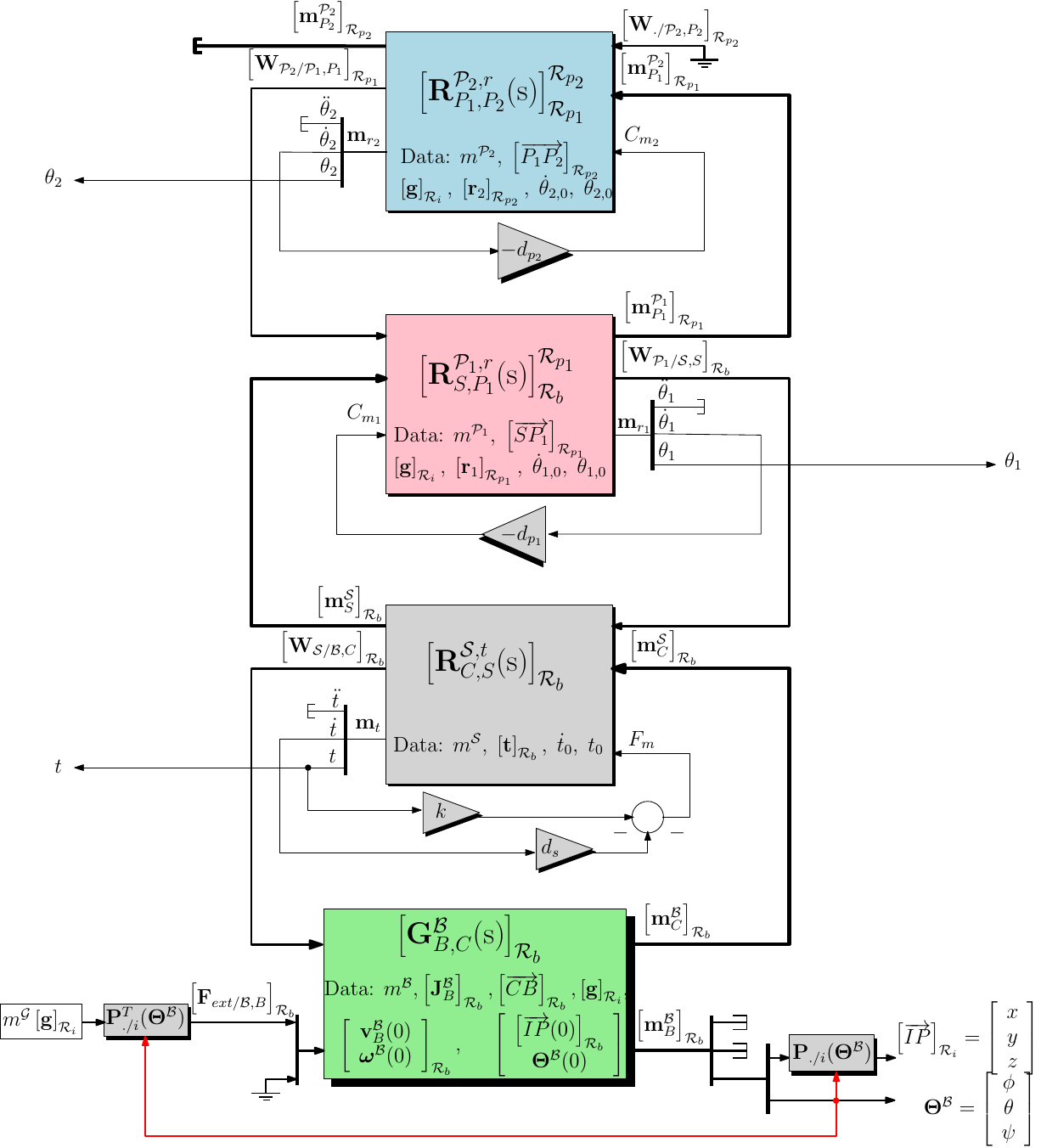}
		\caption{ Block-diagram representation og $\mathcal G$.}
		\label{fig:schema1_blk}
	\end{center}
\end{figure}

\begin{figure}[!ht]
	\begin{center}
	\includegraphics[angle=0, width=\textwidth]{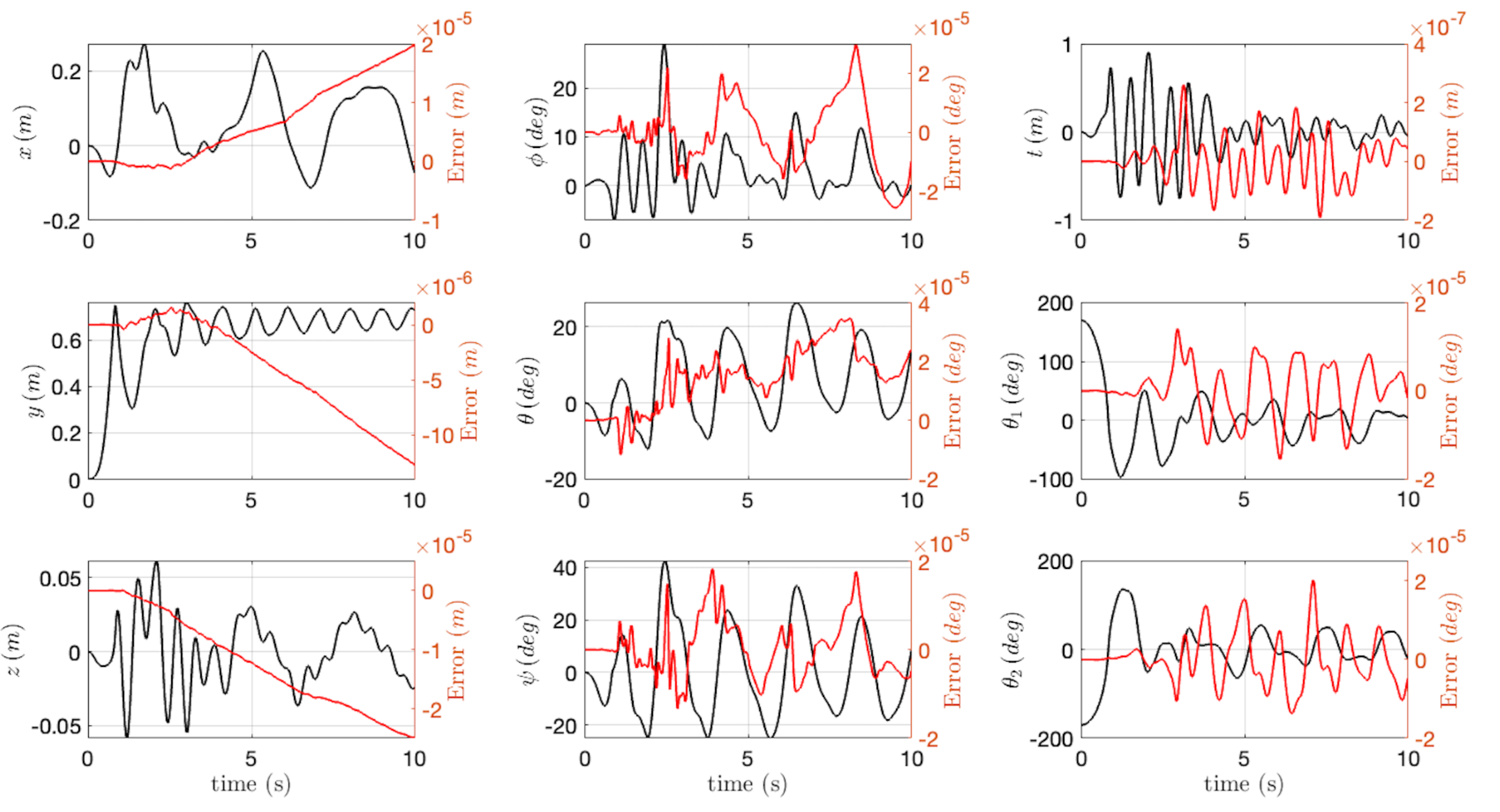}
		\caption{The simulation results.}
		\label{fig:simu}
	\end{center}
\end{figure}

The proposed block-diagram modeling approach is now illustrated on a $9$ d.o.f.  multi-body system and a simulation scenario where the non-linearities are particularly relevant. It is also validated by comparison with the simulation of this system modeled with Simscape-multibody toolbox \cite{mathworks2021simscape}
as depicted in Figure \ref{fig:schema1} thanks to the Mechanical Explorer visualization application. This system $\mathcal G$, working under the Earth gravitational field $\mathbf g$, is composed of a balloon $\mathcal B$ holding a rigid flight chain. At the tip $C$ of this flight chain is connected a slider mass $\mathcal S$ through a prismatic joint working in the plane $(\mathbf x_b,\;\mathbf y_b)$ of the balloon body frame $\mathcal R_b$. The slider mass holds at the point $S$ a double point-mass-pendulum. The $2$ revolute joint axes of the $2$ pendulums $\mathcal P_1$ and $\mathcal P_2$ are aligned with the $\mathbf z_b$-axis. 
The block-diagram, based on the proposed approach, is depicted in Figure \ref{fig:schema1_blk}. Each block is associated with a body and depends only on the dynamic parameters specific to the body and on the initial conditions on its internal d.o.f. The outputs of the model are the positions of the $9$ d.o.f: 
\begin{itemize}
	\item the $3$ components  ($x$, $y$, $z$)  of the position $\left[\overrightarrow{IP}\right]_{\mathcal R_i}$ and the $3$ components ($\phi$, $\theta$, $\psi$)  of the attitude $\boldsymbol \Theta^{\mathcal B}$ of the balloon $\mathcal B$ w.r.t. the inertial frame $\mathcal R_i$
	\item and the $3$  joint  configuration: $p$ for the prismatic joint, $\theta_1$ and $\theta_2$ for the $2$ revolute joints.
\end{itemize}
The balloon is also submitted to an external buoyancy force at its center of mass $B$:
\[
\mathbf F_{ext/\mathcal B,B}=-m^{\mathcal G} \mathbf g
\]
where $m^\mathcal G=m^\mathcal B+ m^\mathcal S+ m^{\mathcal P_1}+m^{\mathcal P_2}$ is the total mass of the system. Finally, an internal stiffness ($k$) and a damping ($d_s$) act inside the prismatic joint while only a damping ($d_{p_1}$ and $d_{p_2}$) acts inside each revolute joint:
\[
F_m=-k  p -d_s \dot p,\quad C_{m_1}=-d_{p_1}\dot\theta_1,\quad C_{m_2}=-d_{p_2}\dot\theta_2\;.
\]
The data and the initial conditions are summarized in Table \ref{tab:na}. The time-domain responses of the $9$ d.o.f. are presented in Figure \ref{fig:simu}. The black plots are the responses obtained with the block-diagram model while the red plots are the errors between these responses and the ones obtained with the Simscape-multibody model. These errors are completely negligible and validate  the proposed model (see also \url{https://youtu.be/e1MVM3VZW7s} for a video of this simulation).


\begin{table}[t!]
	\begin{center}
		\caption{Numaerical application.}
		\label{tab:na}
		\begin{tabular}{c |c |c}
			\toprule
			Body &  Parameters&  Initial conditions \\
			\midrule 
			Balloon	& $m^{\mathcal B}=10\,Kg$, 		$\left[\mathbf J^{\mathcal B}_B\right]_{\mathcal R_b}=\left[\begin{array}{ccc} 5 & 0&  0 \\ 0 & 10 & -7\\ 0 &-7 &10\end{array}\right]\,Kg\,m^2$, 
			&  $\left[\begin{array}{c}\mathbf{v}^{\mathcal B}_B(0)\\ \mbox{\boldmath$\displaystyle\mathbf{\omega}$}^{\mathcal B}(0)\end{array}\right]_{\mathcal{R}_b}=\mathbf 0_{6\times 1}$ \\
			 & $\left[\overrightarrow{CB}\right]_{\mathcal R_b}=\left[\begin{array}{c}
			 	0 \\2 \\ 0
			 \end{array}\right]\,m$, $\left[\mathbf g \right]_{\mathcal R_i}=\left[\begin{array}{c}
			 	0 \\-9.81 \\ 0
			 \end{array}\right]\,m/s^2$ & $\left[\begin{array}{c}\left[\overrightarrow{IP} (0)\right]_{\mathcal{R}_b} \\  \boldsymbol\Theta^{\mathcal B}(0)\end{array}\right]=\mathbf 0_{6\times 1}$ \\ \midrule 
			 Slider & $m^{\mathcal S}=1\,Kg$, $\left[\mathbf t\right]_{\mathcal R_b}\left[\begin{array}{c}
			 	1 \\0 \\ -1
			 \end{array}\right]\,m$ & $t_0=0$, $\dot t_0=0$ \\ \midrule 
			 Pendulum 1 & $m^{\mathcal P_1}=2\,Kg$, $\left[\overrightarrow{SP_1}\right]_{\mathcal R_{p_1}}=\left[\begin{array}{c}
			 	0 \\-1.2 \\ 0
			 \end{array}\right]\,m$,  & $\dot\theta_{1,0}=0$,  \\
			  & $\left[\mathbf r_1\right]_{\mathcal R_{p_1}}=\left[\begin{array}{c}
			  	0 \\0 \\ 1
			  \end{array}\right]\,m$ & $\theta_{1,0}=170\,deg$ \\ \midrule 
			  Pendulum 2 & $m^{\mathcal P_2}=3\,Kg$, $\left[\overrightarrow{P_1P_2}\right]_{\mathcal R_{p_2}}=\left[\begin{array}{c}
			  0 \\-1.6 \\ 0
			  \end{array}\right]\,m$,  & $\dot\theta_{2,0}=0$,  \\
			  & $\left[\mathbf r_2\right]_{\mathcal R_{p_2}}=\left[\begin{array}{c}
			  0 \\0 \\ 1
			  \end{array}\right]\,m$ & $\theta_{2,0}=-170\,deg$ \\
			\bottomrule
		\end{tabular}
	\end{center}
\end{table}

\section{Conclusions and perspectives}
The proposed generalization of the TITOP approach allows a closed-form block-diagram representation of the equations of motions for multi-body systems composed of open or closed kinematic chains or trees of rigid bodies with holomic constraints such as revolute and prismatic joints and thus is fully general. This representation is very compact and user-friendly in comparison with the various recursive method and algorithms  used to solve the equations of motions. A great amount of work is still  required to extend the method to non-linear flexible multi-body systems. This work can be organize into three topics:
\begin{itemize}
	\item the development of an adapted solver. Indeed the comparison, on the proposed example, of the simulation run time was not presented as it was out of the scope of this paper but a dedicated solver to cope with the algebraic loops present in the block-diagram model  is required,
	\item the development of systematic procedures to compute parameter-dependent equilibrium conditions and the corresponding  LPV models. The highly structured block diagram representation should be very adapted and useful to perform this task,
	\item the development of non-linear TITOP model of flexible bodies. The hybrid equations of motion in terms
	of quasi-coordinates as proposed in \cite{Meirovitch2004} seems an interesting approach to capture centrifugal stiffening and softening effects in flexible bodies defined by a finite element model.
\end{itemize}
\bibliography{mybib} 


\begin{thebibliography}{18}
\ifx \bisbn   \undefined \def \bisbn  #1{ISBN #1}\fi
\ifx \binits  \undefined \def \binits#1{#1}\fi
\ifx \bauthor  \undefined \def \bauthor#1{#1}\fi
\ifx \batitle  \undefined \def \batitle#1{#1}\fi
\ifx \bjtitle  \undefined \def \bjtitle#1{#1}\fi
\ifx \bvolume  \undefined \def \bvolume#1{\textbf{#1}}\fi
\ifx \byear  \undefined \def \byear#1{#1}\fi
\ifx \bissue  \undefined \def \bissue#1{#1}\fi
\ifx \bfpage  \undefined \def \bfpage#1{#1}\fi
\ifx \blpage  \undefined \def \blpage #1{#1}\fi
\ifx \burl  \undefined \def \burl#1{\textsf{#1}}\fi
\ifx \doiurl  \undefined \def \doiurl#1{\url{https://doi.org/#1}}\fi
\ifx \betal  \undefined \def \betal{\textit{et al.}}\fi
\ifx \binstitute  \undefined \def \binstitute#1{#1}\fi
\ifx \binstitutionaled  \undefined \def \binstitutionaled#1{#1}\fi
\ifx \bctitle  \undefined \def \bctitle#1{#1}\fi
\ifx \beditor  \undefined \def \beditor#1{#1}\fi
\ifx \bpublisher  \undefined \def \bpublisher#1{#1}\fi
\ifx \bbtitle  \undefined \def \bbtitle#1{#1}\fi
\ifx \bedition  \undefined \def \bedition#1{#1}\fi
\ifx \bseriesno  \undefined \def \bseriesno#1{#1}\fi
\ifx \blocation  \undefined \def \blocation#1{#1}\fi
\ifx \bsertitle  \undefined \def \bsertitle#1{#1}\fi
\ifx \bsnm \undefined \def \bsnm#1{#1}\fi
\ifx \bsuffix \undefined \def \bsuffix#1{#1}\fi
\ifx \bparticle \undefined \def \bparticle#1{#1}\fi
\ifx \barticle \undefined \def \barticle#1{#1}\fi
\bibcommenthead
\ifx \bconfdate \undefined \def \bconfdate #1{#1}\fi
\ifx \botherref \undefined \def \botherref #1{#1}\fi
\ifx \url \undefined \def \url#1{\textsf{#1}}\fi
\ifx \bchapter \undefined \def \bchapter#1{#1}\fi
\ifx \bbook \undefined \def \bbook#1{#1}\fi
\ifx \bcomment \undefined \def \bcomment#1{#1}\fi
\ifx \oauthor \undefined \def \oauthor#1{#1}\fi
\ifx \citeauthoryear \undefined \def \citeauthoryear#1{#1}\fi
\ifx \endbibitem  \undefined \def \endbibitem {}\fi
\ifx \bconflocation  \undefined \def \bconflocation#1{#1}\fi
\ifx \arxivurl  \undefined \def \arxivurl#1{\textsf{#1}}\fi
\csname PreBibitemsHook\endcsname

\bibitem{Schaub}
\begin{bbook}
\bauthor{\bsnm{Schaub}, \binits{H.}},
\bauthor{\bsnm{Junkins}, \binits{J.L.}}:
\bbtitle{Analytical Mechanics of Space Systems}.
\bsertitle{AIAA Education series}.
\bpublisher{American Institute of Aeronautics and Astronautics, Inc.},
\blocation{12700 Sunrise Valley Drive, Reston, VA, 20191-5807}
(\byear{2018})
\end{bbook}
\endbibitem

\bibitem{Hahn}
\begin{bbook}
\bauthor{\bsnm{Hahn}, \binits{H.}}:
\bbtitle{Rigid Body Dynamics of Mechanisms}.
\bpublisher{Springer},
\blocation{Berlin, Germany}
(\byear{2013})
\end{bbook}
\endbibitem

\bibitem{Roy}
\begin{bbook}
\bauthor{\bsnm{Featherstone}, \binits{R.}}:
\bbtitle{Rigid Body Dynamics Algorithms}.
\bpublisher{Springer},
\blocation{Berlin, Germany}
(\byear{2008}).
\doiurl{10.1007/978-1-4899-7560-7}
\end{bbook}
\endbibitem

\bibitem{Shabana}
\begin{bbook}
\bauthor{\bsnm{Shabana}, \binits{A.A.}}:
\bbtitle{Computational Dynamics}.
\bpublisher{John Wiley \& Sons, Ltd},
\blocation{The Atrium, Southern Gate, Chichester, West Sussex, PO19 8SQ, United
  Kingdom}
(\byear{2010}).
\doiurl{0.1002/9780470686850}
\end{bbook}
\endbibitem

\bibitem{Simeon}
\begin{bbook}
\bauthor{\bsnm{Simeon}, \binits{B.}}:
\bbtitle{Computational Flexible Multobody Dynamics - A differential-Algebraic
  Approach}.
\bpublisher{Springer},
\blocation{Berlin, Germany}
(\byear{2013}).
\doiurl{10.1007/978-3-642-35158-7}
\end{bbook}
\endbibitem

\bibitem{oatao16559}
\begin{barticle}
\bauthor{\bsnm{Chebbi}, \binits{J.}},
\bauthor{\bsnm{Dubanchet}, \binits{V.}},
\bauthor{\bsnm{Gonzalez}, \binits{J.A.P.}},
\bauthor{\bsnm{Alazard}, \binits{D.}}:
\batitle{Linear dynamics of flexible multibody systems : a system-based
  approach}.
\bjtitle{Multibody System Dynamics}
\bvolume{41}(\bissue{1}),
\bfpage{75}--\blpage{100}
(\byear{2017}).
\doiurl{10.1007/s11044-016-9559-y}.
\bcomment{Thanks to Springer editor. The definitive version is available at
  http://www.springerlink.com The original PDF of the article can be found at
  Multibody System Dynamics website: http://link.springer.com/journal/11044}
\end{barticle}
\endbibitem

\bibitem{DeNOC}
\begin{barticle}
\bauthor{\bsnm{Saha}, \binits{S.}},
\bauthor{\bsnm{Shah}, \binits{S.}},
\bauthor{\bsnm{Nandihal}, \binits{P.}}:
\batitle{Evolution of the {DeNOC}-based dynamic modelling for multibody
  systems}.
\bjtitle{Mechanical Sciences}
\bvolume{4},
\bfpage{1}--\blpage{20}
(\byear{2013}).
\doiurl{10.5194/ms-4-1-2013}
\end{barticle}
\endbibitem

\bibitem{Vincent}
\begin{botherref}
\oauthor{\bsnm{Dubanchet}, \binits{V.}}:
Modeling and control of a flexible space robot to capture a tumbling debris.
PhD thesis,
Institut Sup{\'e}rieur de l'A{\'e}ronautique et de l'Espace, Polytechnique
  Montr{\'e}al
(2016)
\end{botherref}
\endbibitem

\bibitem{mathworks2021simscape}
\begin{botherref}
\oauthor{\bsnm{{The MathWorks, Inc.}}}:
Simscape Multibody.
[Online; accessed Feb 2021]
(2021).
\url{https://www.mathworks.com/products/simmechanics.html}
\end{botherref}
\endbibitem

\bibitem{AlazardMUBO2023}
\begin{barticle}
\bauthor{\bsnm{Alazard}, \binits{D.}},
\bauthor{\bsnm{Finozzi}, \binits{A.}},
\bauthor{\bsnm{Sanfedino}, \binits{F.}}:
\batitle{Port inversions of parametric two-input two-output port models of
  flexible substructures}.
\bjtitle{Multibody System Dynamics}
\bvolume{57}(\bissue{3}),
\bfpage{365}--\blpage{387}
(\byear{2023}).
\doiurl{10.1007/s11044-023-09883-y}
\end{barticle}
\endbibitem

\bibitem{alazard2020}
\begin{bchapter}
\bauthor{\bsnm{Alazard}, \binits{D.}},
\bauthor{\bsnm{Sanfedino}, \binits{F.}}:
\bctitle{{Satellite dynamics toolbox for preliminary design phase}}.
In: \bbtitle{43rd Annual AAS Guidance and Control Conference, 30 January 2020 -
  5 February 2020 (Brechenridge, United States)},
pp. \bfpage{1461}--\blpage{1472}
(\byear{2020})
\end{bchapter}
\endbibitem

\bibitem{SANFEDINO2022107961}
\begin{barticle}
\bauthor{\bsnm{Sanfedino}, \binits{F.}},
\bauthor{\bsnm{Thiébaud}, \binits{G.}},
\bauthor{\bsnm{Alazard}, \binits{D.}},
\bauthor{\bsnm{Guercio}, \binits{N.}},
\bauthor{\bsnm{Deslaef}, \binits{N.}}:
\batitle{Advances in fine line-of-sight control for large space flexible
  structures}.
\bjtitle{Aerospace Science and Technology}
\bvolume{130},
\bfpage{107961}
(\byear{2022}).
\doiurl{10.1016/j.ast.2022.107961}
\end{barticle}
\endbibitem

\bibitem{Finozzi2022}
\begin{barticle}
\bauthor{\bsnm{Finozzi}, \binits{A.}},
\bauthor{\bsnm{Sanfedino}, \binits{F.}},
\bauthor{\bsnm{Alazard}, \binits{D.}}:
\batitle{Parametric sub-structuring models of large space truss structures for
  structure/control co-design}.
\bjtitle{Mechanical Systems and Signal Processing}
\bvolume{180},
\bfpage{109427}
(\byear{2022}).
\doiurl{10.1016/j.ymssp.2022.109427}
\end{barticle}
\endbibitem

\bibitem{kassarianIEEE}
\begin{barticle}
\bauthor{\bsnm{Kassarian}, \binits{E.}},
\bauthor{\bsnm{Sanfedino}, \binits{F.}},
\bauthor{\bsnm{Alazard}, \binits{D.}},
\bauthor{\bsnm{Chevrier}, \binits{C.-A.}},
\bauthor{\bsnm{Montel}, \binits{J.}}:
\batitle{{Linear Fractional Transformation Modeling of Multibody Dynamics
  Around Parameter-Dependent Equilibrium}}.
\bjtitle{{IEEE Transactions on Control Systems Technology}}
\bvolume{31}(\bissue{1}),
\bfpage{418}--\blpage{425}
(\byear{2023}).
\doiurl{10.1109/TCST.2022.3167610}
\end{barticle}
\endbibitem

\bibitem{rodrigues2024modelinganalysisflexiblespinning}
\begin{botherref}
\oauthor{\bsnm{Rodrigues}, \binits{R.}},
\oauthor{\bsnm{Alazard}, \binits{D.}},
\oauthor{\bsnm{Sanfedino}, \binits{F.}},
\oauthor{\bsnm{Mauriello}, \binits{T.}},
\oauthor{\bsnm{Iannelli}, \binits{P.}}:
Modeling and analysis of a flexible spinning Euler-Bernoulli beam with
  centrifugal stiffening and softening: A Linear Fractional Representation
  approach with application to spinning spacecraft
(2024).
\url{https://arxiv.org/abs/2401.17519}
\end{botherref}
\endbibitem

\bibitem{wiki:NE}
\begin{botherref}
\oauthor{\bsnm{Wikipedia}}:
Newton-Euler equations--- {W}ikipedia{,} The Free Encyclopedia.
[Online; accessed 2-Dec-2024]
(2024).
\url{https://en.wikipedia.org/wiki/Newton-Euler_equations}
\end{botherref}
\endbibitem

\bibitem{meirovitch1970methods}
\begin{bbook}
\bauthor{\bsnm{Meirovitch}, \binits{L.}}:
\bbtitle{Methods of Analytical Dynamics}.
\bsertitle{Advanced engineering series}.
\bpublisher{McGraw-Hill},
\blocation{New York, NY, USA}
(\byear{1970})
\end{bbook}
\endbibitem

\bibitem{Meirovitch2004}
\begin{barticle}
\bauthor{\bsnm{Meirovitch}, \binits{L.}},
\bauthor{\bsnm{Tuzcu}, \binits{I.}}:
\batitle{Unified theory for the dynamics and control of maneuvering flexible
  aircraft}.
\bjtitle{AIAA Journal}
\bvolume{42}(\bissue{4}),
\bfpage{714}--\blpage{727}
(\byear{2004}).
\doiurl{10.2514/1.1489}
\end{barticle}
\endbibitem

\end{thebibliography}


\end{document}